\documentclass[aps,pre,twocolumn,superscriptaddress]{revtex4}
\pdfoutput = 1
\usepackage{times,mathptmx,amsmath,amssymb}
\usepackage{graphicx} %eps figures can be used instead
\usepackage{gensymb}

\usepackage{color}
\newcommand{\amm}[1]{\textcolor{black}{#1}}
\newcommand{\rw}[1]{\textcolor{black}{#1}}
\newcommand{\gka}[1]{\textcolor{black}{#1}}

\begin{document}

\title{Buckling of paramagnetic chains in soft gels}

\author{Shilin Huang} 
\affiliation{
Max Planck Institute for Polymer Research, Ackermannweg 10, D-55128 Mainz, Germany
}

\author{Giorgio Pessot}
\affiliation{
Institut f\"ur Theoretische Physik II: Weiche Materie, Heinrich-Heine-Universit\"at D\"usseldorf, D-40225 D\"usseldorf, Germany
}

\author{Peet Cremer}
\affiliation{
Institut f\"ur Theoretische Physik II: Weiche Materie, Heinrich-Heine-Universit\"at D\"usseldorf, D-40225 D\"usseldorf, Germany
}

\author{Rudolf Weeber}
\affiliation{
Institute for Computational Physics, Universit\"at Stuttgart, D-70569 Stuttgart, Germany
}

\author{Christian Holm} 
\affiliation{
Institute for Computational Physics, Universit\"at Stuttgart, D-70569 Stuttgart, Germany
}

\author{Johannes Nowak} 
\affiliation{
Institute of Fluid Mechanics, Technische Universit\"at Dresden, D-01062, Dresden, Germany
}

\author{Stefan Odenbach} 
\affiliation{
Institute of Fluid Mechanics, Technische Universit\"at Dresden, D-01062, Dresden, Germany
}

\author{Andreas M. Menzel} 
\affiliation{
Institut f\"ur Theoretische Physik II: Weiche Materie, Heinrich-Heine-Universit\"at D\"usseldorf, D-40225 D\"usseldorf, Germany
}

\author{G\"unter K. Auernhammer}
\affiliation{
Max Planck Institute for Polymer Research, Ackermannweg 10, D-55128 Mainz, Germany
}

%Please note that \ast indicates the corresponding author(s) but no footnote text is required.

\begin{abstract}
We study the magneto-elastic coupling behavior of paramagnetic chains in soft polymer gels \amm{exposed to external magnetic fields}. To this end, a laser scanning confocal microscope is used to observe the morphology of the paramagnetic chains \amm{together with} the deformation field \amm{of} the \amm{surrounding gel network}. The paramagnetic chains in soft polymer gels show rich \amm{morphological shape changes} under \amm{oblique magnetic fields, in particular a pronounced buckling deformation}. The \amm{details of the resulting morphological shapes} depend on the length of the chain, \amm{the} strength of \amm{the external} magnetic field, and the modulus of the gel. Based on the observation that the magnetic \amm{chains are} strongly coupled to the \amm{surrounding} polymer network, a simplified model is developed to describe \amm{their buckling behavior}.
\gka{A coarse-grained molecular dynamics simulation model featuring an increased matrix stiffness on the surfaces of the particles leads to morphologies in agreement with the experimentally observed buckling effects.}
\end{abstract}

\maketitle

\section{Introduction}

%Please use \dag to cite the ESI in the main text of the article.
%If you article does not have ESI please remove the the \dag symbol from the title and the above footnotetext.

%additional addresses can be cited as above using the lower-case letters, c, d, e... If all authors are from the same address, no letter is required

%\footnotetext{\ddag~Additional footnotes to the title and authors can be included \emph{e.g.}\ `Present address:' or `These authors contributed equally to this work' as above using the symbols: \ddag, \textsection, and \P. Please place the appropriate symbol next to the author's name and include a \texttt{\textbackslash footnotetext} entry in the the correct place in the list.}

Magneto-responsive hybrid gels (MRGs) have been attracting great attention due to their \amm{tunable} elasticity, swelling propert\amm{ies} and shape that can be remotely controlled by a magnetic field. They have potential applications as soft actuators, artificial muscles, \amm{as well as} sensors \cite{Ilg2013,Snyder2010,zimmermann2006modelling} and can serve as model systems to study the heat transfer in hyperthermal cancer treatment \cite{hergt2006magnetic}. Compared to other stimuli-responsive gels, MRGs have the advantage of fast response, controlled mechanical properties and reversible deformabilities \cite{Szab1998,Abramchuk2006,filipcsei2007magnetic}. A typical MRG consists of a chemically cross-li\amm{n}ked polymer network\amm{,} swollen in a good solvent\amm{,} and embedded magnetic particles \cite{Szab1998,Collin2003}. The size of the magnetic particles can range from $\sim10\;\mathrm{nm}$ to several $\mathrm{\mu}$m \amm{\cite{filipcsei2007magnetic}}.

The origin of the magneto-responsive behavior of MRGs is the magnetic interaction between the magnetic filler particles \amm{as well as their interaction with external magnetic fields} \cite{Faraudo2013,Griffiths1999}. In a uniform magnetic field, paramagnetic particles can be polarized and act as approximate magnetic dipoles. \amm{Depending on their mutual azimuthal configuration, the dipolar interactions can be either attractive or repulsive.} In a liquid carrier, the dipolar interaction drives the magnetic particles to form chains and columns \cite{klapp2002spontaneous,klapp2005dipolar,Gajula2010,deVicente2011} aligning in the direction of the magnetic field. However, in a polymer gel, the particles cannot change their position freely. Instead, \amm{relative displacements of the particles, induced e.g.\ by changes in the magnetic interactions,} lead to \amm{opposing} deformations \amm{of} the polymer network\amm{. As a result, the magnetic interactions can induce changes in} the modulus of the gel \cite{Auernhammer2006,filipcsei2007magnetic}. This magneto-elastic effect is well known to be related to the spatial distribution of the magnetic particles \cite{wood2011modeling,ivaneyko2012effects,Han2013,Pessot2014,Menzel2015,stolbov11a}.  For example, the modulus of anisotropic materials \gka{that contain aligned chain-like aggregates of magnetic filler particles \cite{Auernhammer2006,Filipcsei2007,Gunther2012,borbath2012xmuct}} can be significantly enhanced when \amm{an external} magnetic field is applied \amm{along} the chain direction \cite{filipcsei2007magnetic}. The anisotropic arrangement of particles also dominates the anisotropic magnetostriction effects \cite{Guan2008,Danas2012,zubarev2013effect}. 

\amm{Different} theoretical routes have been pursued to investigate the magneto-elastic \amm{effects} of MRGs\amm{:} macroscopic continuum mechanics approaches \amm{\cite{jarkova2003hydrodynamics,bohlius2004macroscopic}, mesoscopic modeling \cite{wood2011modeling,ivaneyko2012effects,Han2013,Pessot2014}, and more microscopic approaches \cite{Weeber2012,weeber2015ferrogels,ryzhkov2015coarse} that resolve individual polymer chains. Theoretical routes to connect and unify these different levels of description have recently been proposed \cite{menzel2014bridging,Ivaneyko2014,pessot2015towards}}. The authors \amm{of Ref.~\citenum{Ivaneyko2014}} show \amm{how the interplay between the mesoscopic} particle distribution and the macroscopic shape of the sample \amm{affects} the magneto-elastic effect. \amm{In addition to} these factors, recent experiments and computer simulations also point out that \amm{a direct} coupling between the magnetic particles and \amm{attached polymer chains can play} another important role \cite{Ilg2013,Weeber2012,Roeder2014, frickel2009functional,frickel2011magneto,messing2011cobalt,weeber2015ferrogels}.

An experimental model system showing a well-defined particle distribution and a measurable magneto-elastic effect can help to understand the magneto-elastic behavior of MRGs at different length scales. \amm{Projected into a two-dimensional plane, the} distribution of magnetic particles in thin diluted MRGs can be detected using optical \amm{microscopy} or light scattering methods \cite{Auernhammer2006,Csetneki2006}. By combining these techniques with magnetic or mechanical devices, it is possible to observe the particle rearrangement when the MRG sample is exposed to a magnetic field or mechanical stimuli \cite{Auernhammer2006,An2014}. For three-dimensional (3D) characterization, X-ray micro-tomography has been used \cite{Gunther2012}. Here we introduce another 3D imaging technique \amm{--} laser scanning confocal microscopy (LSCM). Compared to normal optical microscopy, LSCM is able to observe 3D structures and it has a better resolution \cite{Minsky1988}. Compared to X-ray micro-tomography, LSCM is faster in obtaining a 3D image and easier to combine with other techniques for real-time investigation \cite{roth2011,Roth2012}.

We use LSCM to study the magneto-elastic effects of paramagnetic chains in soft \amm{gels. As} a result, we find that the paramagnetic chains in soft gels (elastic modulus $<2\;\mathrm{Pa}$) under an oblique magnetic field show rich morphologies. Depending on the length of the chain, modulus of the gel and strength of \amm{an external} magnetic field, the chains can rotate, bend and buckle. The deformation field in the polymer network around the deformed paramagnetic chains can also be tracked. The result confirms that the chains are strongly coupled to the polymer network. A simplified model is developed to understand the magnetically induced buckling behavior of the paramagnetic chains in soft gels. \amm{In addition to} serving as a model experimental system for studying the magneto-elastic effect of MRGs, our \amm{approach} may also provide a new microrheological technique to probe the mechanical property of a soft gel \cite{Wilhelm2008}. \amm{Furthermore, our results may} be interesting to biological scientists who study how magnetosome chains interact with the surrounding cytoskeletal network in magnetotactic bacteria \cite{Kornig2014}.

\section{Materials and Methods}
The elastic network was obtained by hydrosilation of a difunctional vinyl-terminated polydimethylsiloxane (\gka{vinyl-terminated PDMS, }DMS-V25, Gelest Inc.) prepolymer with a SiH-containing cross-linker (PDMS, HMS-151, Gelest Inc.). Platinum(0)-1,3-divinyl-1,1,3,3-tetramethyldisiloxane complex (Alfa Aesar) was used as a catalyst. A low-molecular-weight trimethylsiloxy terminated PDMS (770 g/mol, Alfa Aesar GmbH \& \amm{Co.\ }KG, in the following ``PDMS 770'') served as a solvent that carried the polymer network and the paramagnetic particles. The paramagnetic particles were purchased from microParticles GmbH. They were labeled with fluorophores (visible in LSCM). The materials consist of porous polystyrene spheres. Within the pores, nanoparticulate iron oxide was distributed rendering the particles superparamagnetic. To prevent iron oxide leaching, the particles had a polymeric sealing that also held the fluorophores. The particles had a diameter of $1.48\pm0.13\ \mathrm{{\mu}m}$ %(Fig.~S1a). 
\cite{suppl}. 
We measured the magnetization curve %(Fig.~S1b) 
\cite{suppl} of the paramagnetic particles by a vibrating sample magnetometer (VSM, Lake Shore 7407). We found about 20\% deviations in the magnetic properties of the magnetic particles (e.g., magnetic moment \cite{suppl}). %Fig.~S2). 
In order to observe the deformation field in the polymer network, we used fluorescently labeled silica particles as tracers. They had a diameter of $830\pm50\ \mathrm{nm}$ and the surface was modified with N,N-dimethyl-N-octadel-3-amino-propyltrimethoxy\amm{silyl}chloride.

The paramagnetic particles were dried in a vacuum oven at room temperature overnight before they were dispersed into PDMS 770. The prepolymer mixture was prepared with 9.1 wt\% vinyl-terminated PDMS and 90.9 wt\% SiH-containing cross-linker. The \amm{prepolymer mixture} (2.86 wt\%) was dissolved in PDMS \amm{770, which} contained the paramagnetic particles. Finally, by adding PDMS 770, which carried the catalyst, the concentration of the prepolymer mixture in the sol solution was carefully adjusted in the range \amm{from} 2.74 wt\% to 2.78 wt\%. This concentration range guaranteed the formation of soft gels with an elastic modulus lower than 10~Pa. In the sol solution, the catalyst concentration was 0.17 wt\%, and the concentration of magnetic particles was 0.09 wt\%. The sol solution was agitated at 2500 r/min with a Reax Control (Heidolph, Schwabach, Germany) for 2 min for homogenization, followed by ultrasonication (2~\amm{min, }Transsonic 460/H, Elma) to disperse the magnetic particles. Then the sol solution was introduced into a thin sample cell ($\sim 160\;\mathrm{\mu m}$ thick and $\sim 1\;\mathrm{cm}$ wide) %, see Fig.~\ref{Fig.1}a) 
by capillary forces. The sample cells consisted of two \amm{No.~1} standard coverslips, separated by $\sim 160\ \mathrm{\mu m}$ spacers. After sealing with two-component glue, the cells that contained the sol were exposed to a $100.8 \pm 0.5\;\mathrm{mT}$ magnetic field. The paramagnetic particles aligned into chains along the direction of the applied magnetic field while the prepolymer was crosslinking. A visible reaction of the prepolymer occurred within 10 min, and the rheological measurements showed that it took about 40 min to form a gel. Due to the low concentration of magnetic particles, the magnetic chains in the gel were well separated ($>  30\ \mathrm{\mu m}$). The length of the chains \amm{varied} from a single particle up to about 170 particles. In some samples, 3 wt\% silica particles were added as tracers. We stored the samples at ambient temperature for at least two weeks before testing. The modulus of the gel in the sample cells was characterized using microrheological techniques \cite{Mason2000,suppl}. %, see Fig.~S3.

A home-built LSCM setup was used to observe the chain structure in the gel \cite{roth2011,Roth2012}. We were able to analyze a sample of thickness of about $150\;\mathrm{\mu m}$. A homogeneous magnetic field was attained by building Halbach magnetic arrays near the sample stage of the LSCM \cite{Raich2004}. A 32-magnet array (Fig.~\ref{Fig.1}a) was used to change the field direction while keeping the field strength constant ($216.4\pm1.1\;\mathrm{mT}$ \cite{suppl}). % Fig.~S4a). 
Another 4-magnet Halbach array \cite{suppl} %(Fig.~S4c) 
was used to change the field strength (up to $100.8\pm0.5\;\mathrm{mT}$). The magnetic field was measured by a Lake Shore Model 425 Gaussmeter with a transverse probe.

\section{Results}
\begin{figure}[tb]
\centering
  \includegraphics[width=8.5cm]{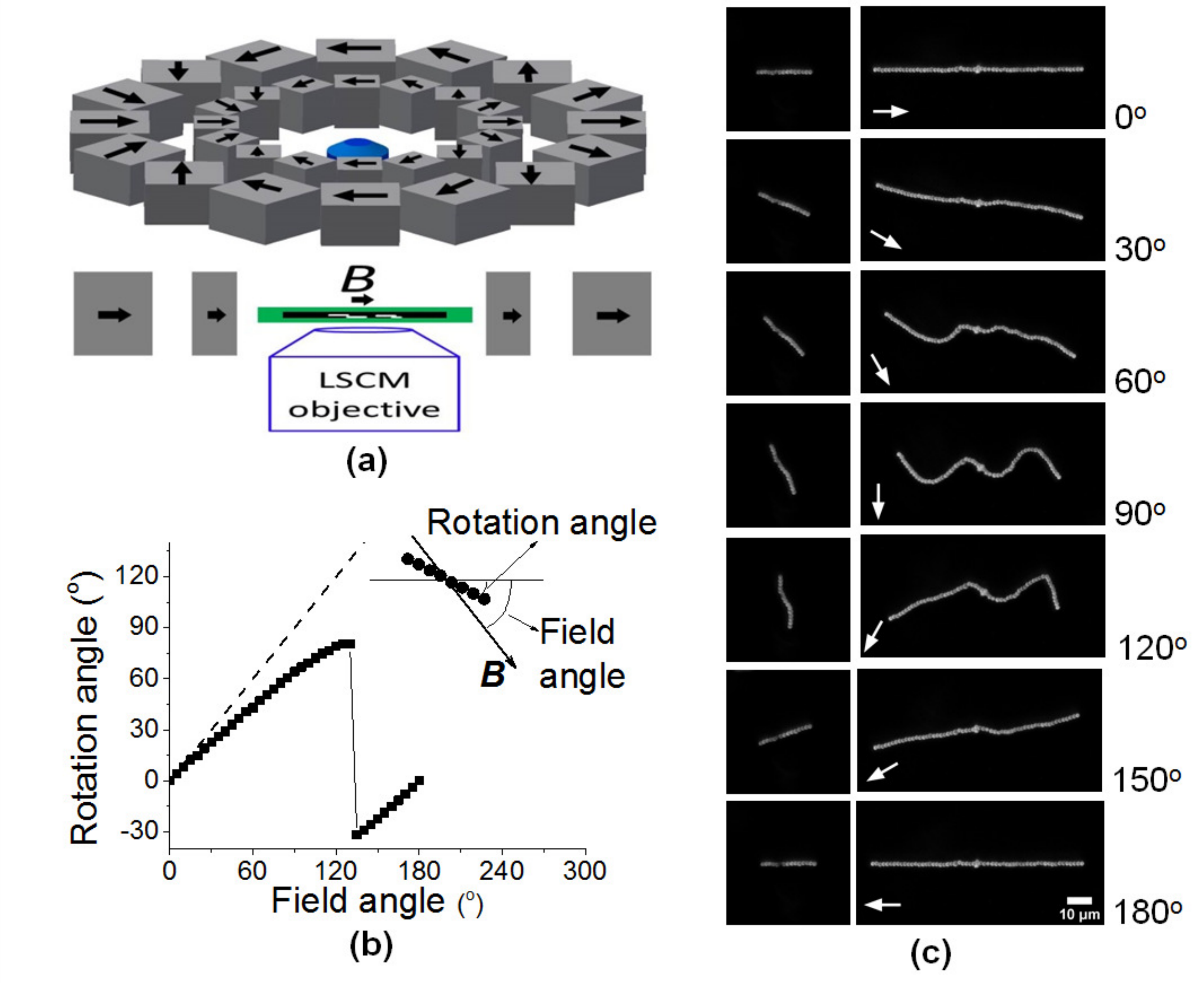}
  \caption{(a) Laser scanning confocal microscopy (LSCM) was used to observe the morphologies of the paramagnetic chains in the soft gels. The Halbach magnetic array provided a homogeneous magnetic field (here $\amm{B} = 216.4\pm1.1\;\mathrm{mT}$). This array could be rotated to change the orientation of the magnetic field. (b) The orientation of the magnetic field $\pmb{B}$ was successively increased from $0\degree$ to $180\degree$ in 36 steps (square points). \amm{A} magnetic chain \amm{of} 15 particles rotated to follow the magnetic field, but the rotation angle was smaller than the orientation \amm{angle} of $\pmb{B}$ (dashed line). (c) Morphologies of magnetic chains in a soft gel change when the orientation \amm{angle} of $\pmb{B}$ increased. The scale bar is 10 $\mu m$. The gel in (b) and (c) had a storage modulus $G'$ of $0.25\pm0.06\;\mathrm{Pa}$.}
  \label{Fig.1}
\end{figure}

In the absence of a magnetic field, the paramagnetic chains in a soft gel kept the aligned morphologies \cite{suppl}. When a magnetic field ($216.4\pm1.1\;\mathrm{mT}$) was applied in the direction parallel to the chains (Fig.~\ref{Fig.1}c, images for $0\degree$), the paramagnetic chains still aligned with the original chain direction (horizontal). We changed the direction of the magnetic field step-by-step ($5\degree$) in the clockwise \amm{direction ($\sim$1} min between steps, quasi-static). We also define the orientation of the magnetic field $\pmb{B}$ as the angle included between the magnetic field and the initial chain direction (see Fig.~\ref{Fig.1}b). The left images of Fig.~\ref{Fig.1}c show a short chain with 15 particles in a gel \amm{of} storage modulus $G'$ of $0.25\pm0.06\;\mathrm{Pa}$. The chain rotated to follow the magnetic field. However, the rotation angle of the chain is smaller than the orientation angle of the magnetic field (Fig.~\ref{Fig.1}b). This difference increased until the orientation of $\pmb{B}$ reaches $135\degree$, where the chain flipped backward and had a negative \amm{angle. The} chain again became parallel to the field when the orientation of $\pmb{B}$ increased to $180\degree$. The morphology of the chain was the same at orientations of the magnetic field of $0\degree$ and $180\degree$ because of the superparamagnetic nature of the particles. Note that the chain was not straight at the intermediate angles (e.g., images for $60\degree$, $90\degree$ and $120\degree$). Instead it bended.

The \amm{images on the right-hand side of} Fig.~\ref{Fig.1}c show a longer chain with 59 particles in the same gel. When the orientation of $\pmb{B}$ was $30\degree$, the chain rotated and bended, with its two ends tending to point in the direction of the magnetic field. When the orientation of $\pmb{B}$ was $60\degree$, the chain wrinkled and started to buckle. A sinusoidal-shape buckling morphology was observed when the magnetic field \amm{was} perpendicular to the original chain (orientation of the magnetic field \amm{of} $90\degree$). When the orientation of $\pmb{B}$ increased from $90\degree$ to $120\degree$, the left part of the chain flipped downward in order to follow the magnetic field. The right part flipped upward when the orientation of $\pmb{B}$ increased from $120\degree$ to $150\degree$. Finally, when the field direction was again parallel to the original chain direction (orientation of the magnetic field \amm{of} $180\degree$), the chain became straight. The same rotation/buckling morphologies as in Fig.~\ref{Fig.1}c could be observed when increasing the orientation of $\pmb{B}$ from $180\degree$ to $360\degree$.

\begin{figure}[tb]
\centering
  \includegraphics[height=5.5cm]{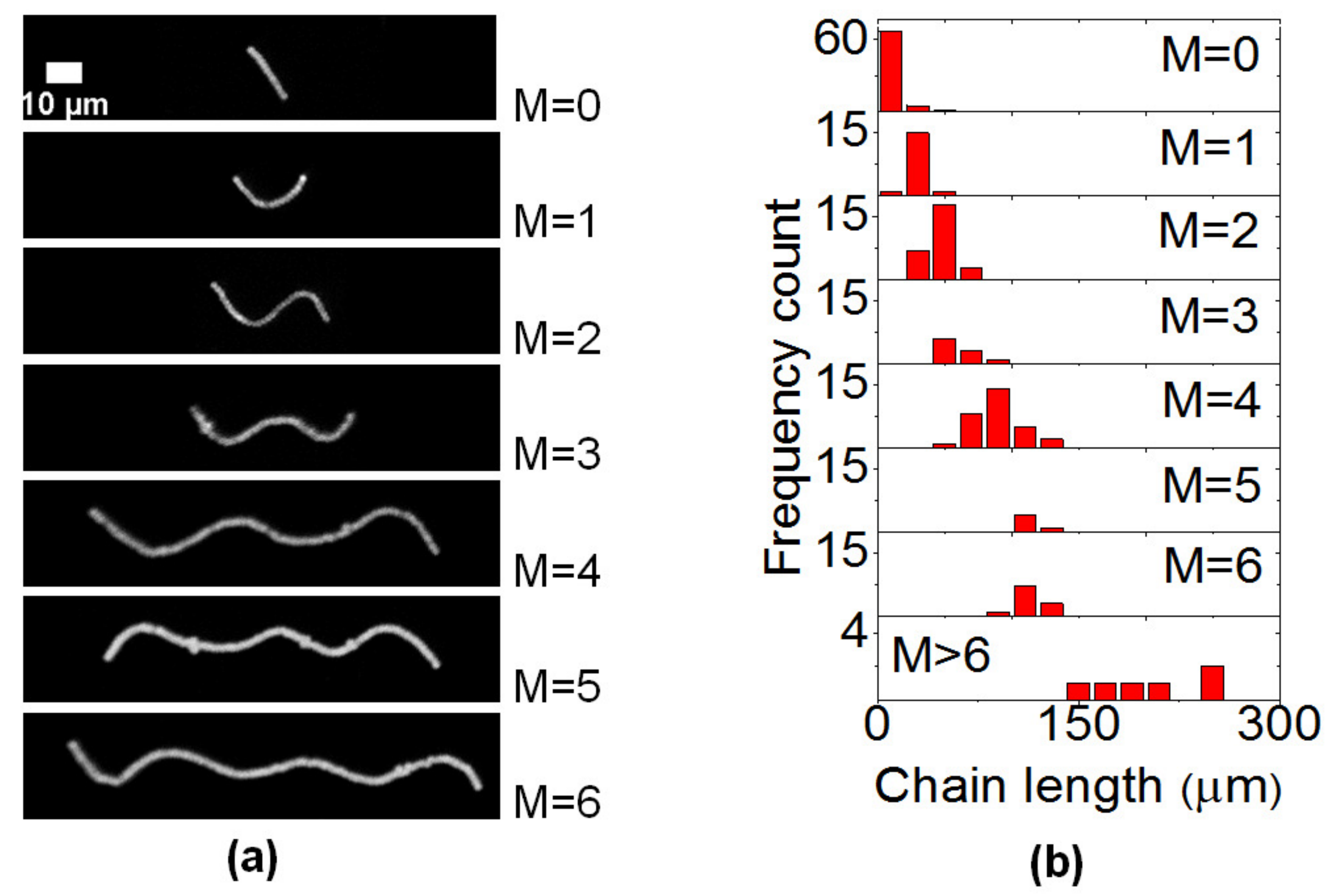}
  \caption{(a) Different morphologies of paramagnetic chains in a soft gel ($G'=0.25\pm0.06\;\mathrm{Pa}$) under a perpendicular magnetic field ($100.8\pm0.5\;\mathrm{mT}$). The original chain direction was horizontal, and the applied magnetic field was vertical. The scale bar is $10\;\mathrm{\mu m}$. (b) Frequency count of different buckling morphologies in the same sample. $M$ is the number of half waves.}
  \label{Fig.2}
\end{figure}

We also directly applied a perpendicular magnetic field to the paramagnetic chains in the soft gels. The paramagnetic chains showed different buckling morphologies (Fig.~\ref{Fig.2}a) depending on the chain length. Fig.~\ref{Fig.2}b gives frequency counts of the different morphologies in the same sample ($G'=0.25\pm0.06\;\mathrm{Pa}$) under a magnetic field of $100.8\pm0.5\;\mathrm{mT}$. \gka{In total 180 chains were observed.} Longer chains tended to buckle with a higher number of half waves. Moreover, the distributions overlapped, implying that the paramagnetic chains with the same length could have different morphologies under the perpendicular magnetic field.

These buckling morphologies are reminiscent of the buckling of paramagnetic chains in a liquid medium under a perpendicular magnetic field \cite{Goubault2003,Shcherbakov2004}. The most stable morphology in the latter system was a straight chain aligning along the magnetic field direction. However, in our system this morphology was not observed. Even the short chains showed a rotation angle smaller than the orientation  of the magnetic field (e.g., Fig.~\ref{Fig.1}b). The major difference between our experiments and Refs.~\citenum{Goubault2003} and \citenum{Shcherbakov2004} was the nature of the surrounding medium. In our system, the polymer network around the paramagnetic chains impeded the rotation of the chains into the magnetic field direction (a more detailed discussion will be given below).

\begin{figure}[t]
\centering
  \includegraphics[width=8.2cm]{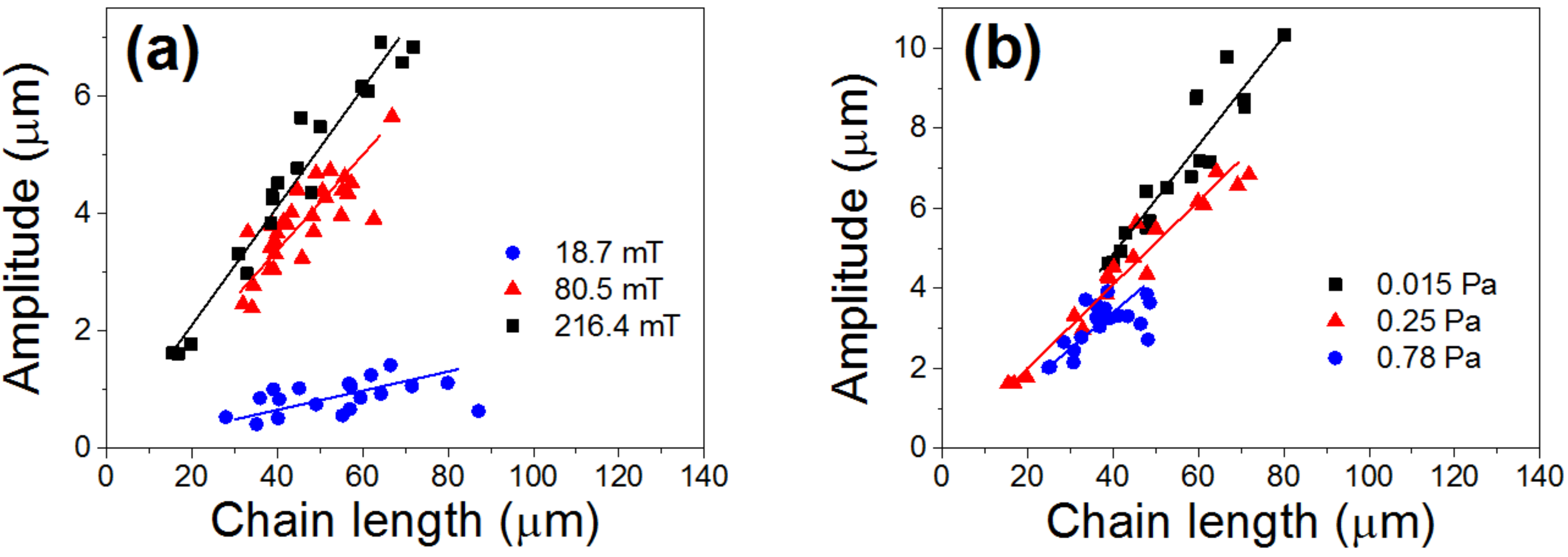}
  \caption{Influence of chain length, strength of magnetic field and elastic modulus of \amm{the} gel matrix on the amplitude of the S-shaped chains, observed when the magnetic field is applied perpendicularly to the initial chain orientation. (a) The elastic modulus of the gel was $0.25\pm0.06\;\mathrm{Pa}$, and the magnetic field strengths were $216.4\pm1.1\;\mathrm{mT}$ (black squares), $80.5\pm0.4\;\mathrm{mT}$ (red triangles) and $18.7\pm0.1\;\mathrm{mT}$ (blue circles), respectively. (b) The magnetic field strength was $216.4\pm1.1$ mT and the elastic moduli of the gel were $0.015\pm0.005\;\mathrm{Pa}$ (black squares), $0.25\pm0.06\;\mathrm{Pa}$ (red triangles) and $0.78\pm0.22\;\mathrm{Pa}$ (blue circles), respectively. The solid lines represent linear fits and are included as guides to the eye.}
  \label{Fig.3}
\end{figure}

We used IMAGEJ software (NIH \cite{Imagej}) to extract the skeletons of the chains which have 2 half waves (S-shaped). The amplitude of \amm{deflection or deformation of} different chains was \amm{quantified by} the square root of the mean square displacement, i.e.\ \gka{$ \mathrm{Amplitude} =(\langle{y^2}\rangle-\langle{y}\rangle^2)^{1/2}$}, where $y$ measures the particle displacement along the field direction. The results are shown in Fig.~\ref{Fig.3}. The amplitude increased with increasing chain length. At the same chain length, the amplitude \amm{tends to increase} with increasing magnetic field strength (Fig.~\ref{Fig.3}a; an example is also given in Fig.~\ref{Fig.4}a) or with decreasing gel modulus (Fig.~\ref{Fig.3}b).

The modulus dependence of the amplitude demonstrated that the polymer network around the paramagnetic chains impeded the \amm{chain deformations}. Therefore, the deformation field within the polymer network \amm{plays an important role} to understand the buckling of the chains. We thus added tracer particles into the gel matrix, and used their \amm{trajectories} to record the deformation field around the paramagnetic chains. As shown in Fig.~\ref{Fig.4}a, a linear paramagnetic chain buckled and formed an \amm{S shape} in a perpendicular magnetic field. The amplitude increased with increasing field strength, while the contour length of the chain remained constant. The chain \amm{extension decreased} along the original chain direction (horizontal direction) and \amm{increased} along the perpendicular direction. Simultaneously, the polymer network around the chain followed the deformation (Fig.~\ref{Fig.4}b) of the paramagnetic chain, both in the transverse and longitudinal directions. This confirmed that the paramagnetic chain is strongly coupled to the polymer network. Within our experimental resolution, the chain seemed to \amm{have} a rigid non-slip contact to the surrounding network.

\section{Modeling}

We now turn to a qualitative description of the situation in the framework of a reduced minimal model.
Theoretically capturing in its full breadth the problem of displacing rigid magnetic inclusions in an
 elastic matrix is a task of high complexity and enormous computational effort \cite{spieler2013xfem}.
We do not pursue this route in the following. Instead, we reduce our characterization to
 a phenomenological description in terms of the shape of the magnetic chain only.
This is possible if the dominant modes of deformation of the surrounding matrix are reflected by the deformational modes of the magnetic chain. 

\begin{figure}[tb]
\centering
  \includegraphics[width=8.3cm]{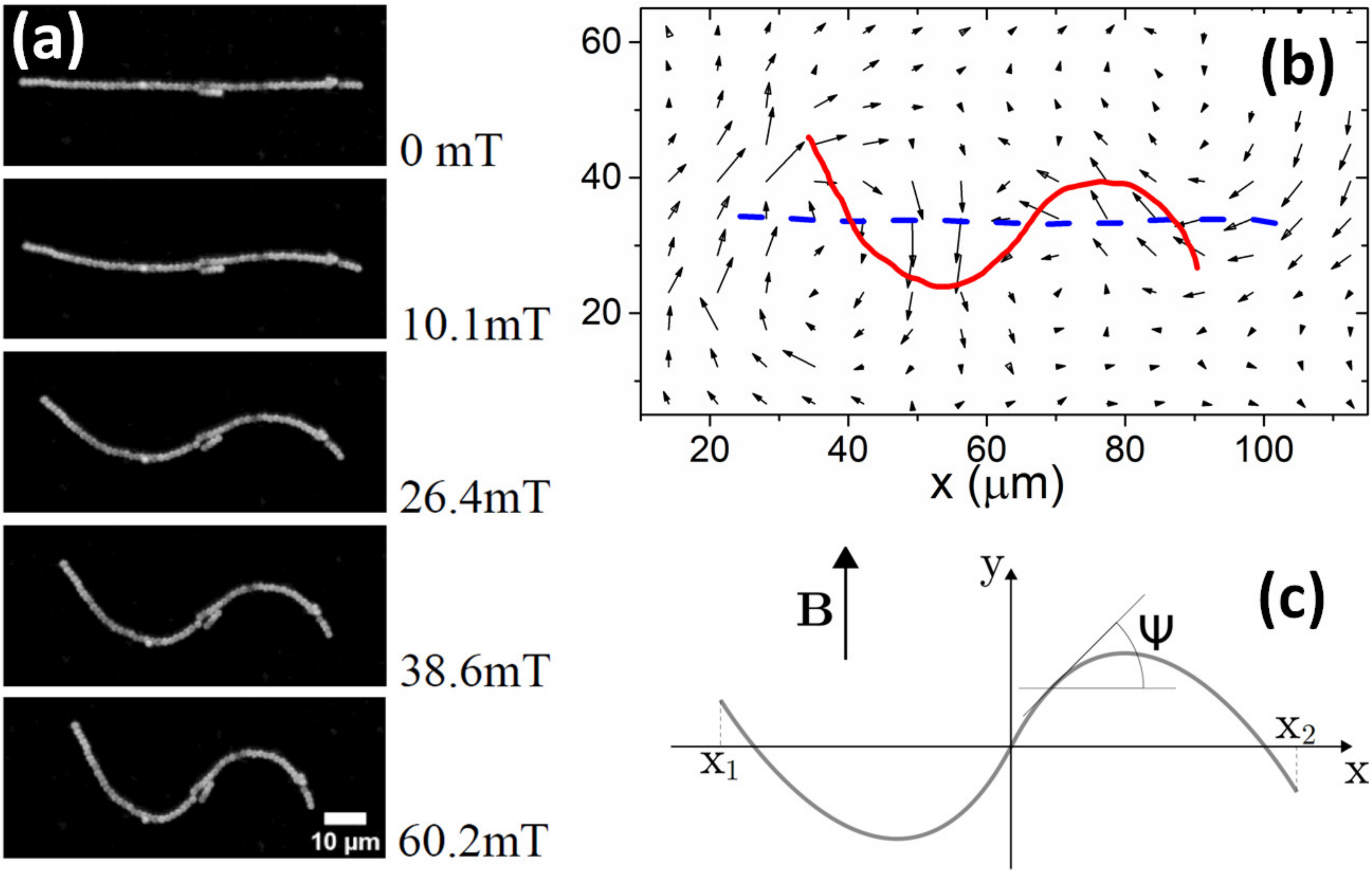}
  \caption{(a) Influence of the magnetic field strength on a buckling chain. From top to bottom, the magnetic field strengths were $0\;\mathrm{mT}$, $10.1\pm0.1\;\mathrm{mT}$, $26.4\pm0.1\;\mathrm{mT}$, $38.6\pm0.2\;\mathrm{mT}$ and $60.2\pm0.3\;\mathrm{mT}$, respectively. The modulus of this gel was about 0.01 Pa. (b) \amm{Tracer particles were inserted into the gel matrix of the sample. Tracking these embedded tracer particles, the deformation field in the gel matrix was determined} (\amm{the} Particle Tracker plug-in developed on IMAGEJ software \amm{was used for this purpose}) \cite{Sbalzarini2005}. The red solid line \amm{represents} the skeleton of the magnetic chain shown in (a) \amm{for a field strength of $60.2\pm0.3\;\mathrm{mT}$}, and the dashed blue line indicates the original chain shape. (c) We modeled the paramagnetic chain in \amm{the} elastic gel as \amm{a continuous object} uniformly \amm{carrying dipolar magnetic moments}. Without the magnetic field, the straight \amm{chain} was \amm{oriented along the $x$-axis}. Under a perpendicular magnetic field $\pmb{B}$ (\amm{oriented along the $y$-axis}), the \amm{magnetic chain} deformed. The surrounding polymer network impeded the \amm{chain deformation.} }
  \label{Fig.4}
\end{figure}

Below, we assume $N$ identical particles on the chain. In the undeformed state,
 the straight chain is located on the $x$-axis of our coordinate frame.
The contour line of the deformed chain running through the particle centers is parameterized as $y(x)$, see Fig.~\ref{Fig.4}c. 

\subsection{Magnetic Energy}
First, concerning the magnetic energy along the chain, we assume dipolar magnetic interactions
 between the particles.
In the perpendicular geometry, the external magnetic field approximately aligns all dipoles along the $y$-axis.
For simplicity, we only include nearest-neighbor magnetic interactions.
In an infinite straight chain, this would result in an error given by a factor of $\zeta(3)\approx1.2$,
 where $\zeta$ is the Riemann zeta function \cite{annunziata2013hardening,menzel2014bridging,prokopieva09a}.
Within our qualitative approach this represents a tolerable error.
Replacing the magnetic interaction energy between the discrete magnetic particles by a continuous
 line integral and shifting the path of integration from the contour line of the chain to the $x$-axis, we obtain the magnetic interaction energy \cite{suppl}
\begin{equation}\label{Emagnetic}
E_{magn} = W\int_{x_1}^{x_2}\frac{1}{\amm{\sqrt{1+y'(x)^2}}} \:dx, 
\end{equation}
where $x_1$ and $x_2$ label the end points of the chain.
The prefactor $W$ has the dimension of energy per unit length and is given by \cite{suppl}
\begin{equation}
W \approx \frac{3\mu_0 m^2}{4\pi d^4},
\end{equation}
where $\mu_0$ is the vacuum magnetic permeability, $m$ the magnetic moment of a single particle,
 and $d$ its diameter.

\subsection{Elastic Bending Energy}
Next, we need to include terms that provide a measure for the magnitude of the elastic deformation energy.
To estimate the importance of different modes of the elastic matrix deformation,
 we analyze the experimentally determined displacement field around the distorted chain shown in Fig.~\ref{Fig.4}b.
For this purpose, we model the continuous matrix by a discretized spring network \cite{Pessot2014,tarama2014tunable}.
 Network nodes are set at the positions where the displacement field was tracked
 experimentally by tracer particles.
The nodes are then connected by elastic springs.
After that, we determine the normal modes of deformation of this network \cite{tarama2014tunable}.
Finally, we can decompose the experimentally observed deformation field in Fig.~\ref{Fig.4}b into these normal modes. Occupation numbers
 $\phi_n$ give the contribution of the $n$th mode to the overall deformation. 

\begin{figure}
\centering
\includegraphics[width=8.6cm]{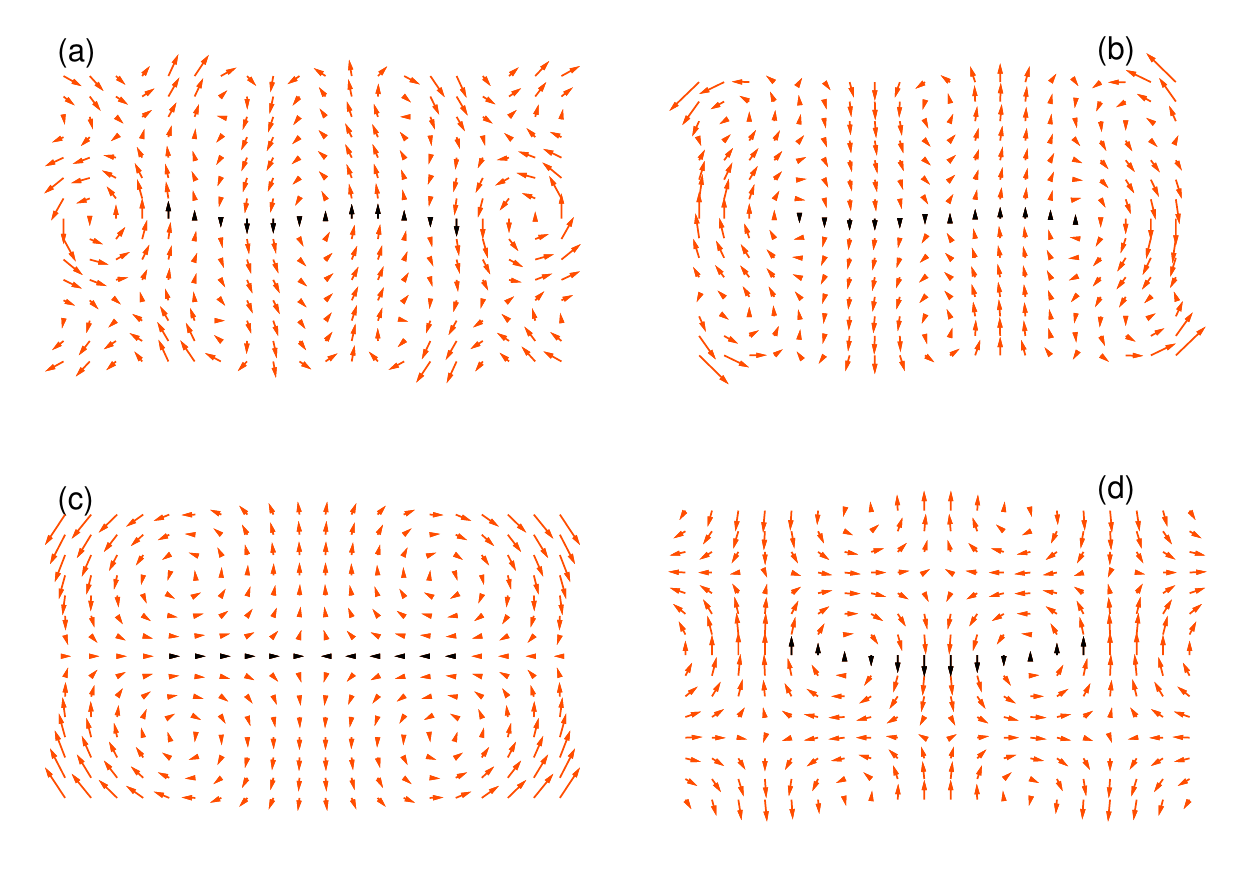}
\caption{The four most occupied normal modes of the deformation
 in Fig.~\ref{Fig.4}b after projection to an elastic spring network,
ordered by decreasing magnitude of contribution to the overall deformation.
The normal modes (a), (b), and (d) are of an ``oscillatory'' type, whereas mode (c) represents
 a longitudinal contraction.
Corresponding relative weights of the modes are $\phi_{(a)}^2=0.095$, $\phi_{(b)}^2=0.057$,
 $\phi_{(c)}^2=0.055$, and $\phi_{(d)}^2=0.051$, where we normalized the sum of the weights over all modes to unity.
For better visualization, the overall amplitudes are rescaled as against the actual weights.
The matrix region in close vicinity of the chain is indicated by black arrows.}
\label{modes}
\end{figure}

The four most occupied modes are shown in Fig.~\ref{modes}.
We find a major contribution of ``oscillatory'' modes, i.e.~alternating
 up and down displacements along the central horizontal axis.
Such oscillatory displacements of the matrix are induced or at least reflected by corresponding
 oscillatory displacements of the chain, see Fig.~\ref{Fig.4}b. A bending term of the form \cite{suppl}
\begin{equation}\label{Ebend}
E_{bend} = C_b\int_{x_1}^{x_2}\frac{[y''(x)]^2}{\amm{\left[1+y'(x)^2\right]^{5/2}}} \:dx
\end{equation}
becomes nonzero when such deformational modes occur and is therefore taken as a measure for their energetic
 contribution.
In addition to that, we have experimental evidence that the chain itself shows a certain
 amount of bending rigidity \cite{suppl}, possibly due to the adsorption of polymer chains on the surfaces
 of the magnetic particles. \amm{Similar indication follows from two-dimensional model simulations, see below.} 

\subsection{Elastic Displacement Energy}\label{modelingdisplsusbs}
The bending term does not energetically penalize rotations of a straight chain, see Fig.~\ref{Fig.2}a for $M=0$.
Yet, such rotations cost energy.
Boundaries of the block of material are fixed, %in its orientation at the boundaries and such straight chains 
therefore any displacement of an inclusion induces a distortion of the surrounding gel matrix.
%This is true for any displacement of an inclusion. 
We model this effect by a contribution \cite{suppl}
\begin{equation}\label{Edisplacement}
E_{displ} = C_d\int_{x_1}^{x_2}\left[y(x)\right]^2\amm{\left[1+y'(x)^2\right]^{3/2}} \:dx.
\end{equation}
This term %,in contrast to a possible contribution $\propto [y'(x)]^2$, 
increasingly disfavors the rotations of longer straight chains, which reflects the experimental observations \cite{suppl}.

\amm{Moreover, in Fig.~\ref{modes}c the third dominating mode of the matrix deformation corresponds to a contraction along the chain direction and an expansion perpendicular to it. 
We conjecture that this should be the dominating mode in the deformational far-field, yet this hypothesis needs further investigation. 
It is induced by chain deflections in $y$-direction, which imply a shrinking extension in $x$-direction (experimentally we observe that the chain length is conserved under deformations and that the individual magnetic particles remain in close contact). We simultaneously use $E_{displ}$ to represent the energetic contribution of this type of underlying matrix deformation.}

%\subsection{Elastic Contraction Energy}
%Finally, in Fig.~\ref{Fig.4}c the third dominating mode corresponds to a contraction along the chain direction and an expansion perpendicular to it. We conjecture that this should be the dominating mode in the deformational far-field, yet this hypothesis needs further investigation. A measure for the contraction is the relative approach between the two end points $x_1$ and $x_2$. This can be calculated from the constraint of conserved total contour length $L$ of the chain (experimentally, we observe that the individual magnetic particles of the chain remain in contact), 
%\begin{equation}\label{eqL}
%\int_{x_1}^{x_2}\sqrt{1+[y'(x)]^2} \:dx = L. 
%\end{equation}
%To lowest order on the coarsest scale, we thus add an energetic penalty 
%\begin{equation}\label{Econtraction}
%E_{contr} = C_c\frac{[L-(x_2-x_1)]^2}{L}. 
%\end{equation}

\subsection{Energetic evaluation}
We now consider the resulting phenomenological model energy
 $E_{tot}=E_{magn}+E_{bend}+E_{displ}$. %+E_{contr}$. 
First, we only address the bulk terms of the energetic expressions.
Minimizing them with respect to the functional form of $y(x)$ and linearizing the final expression,
 one obtains, above a certain threshold of the magnetic field, wave-like oscillatory deformations \cite{suppl}. 
This is in agreement with the observation of the wrinkles at onset in Fig.~\ref{Fig.1}c
 and the final oscillatory shapes in the inner part of the longer chains in Fig.~\ref{Fig.1}a. 
%
%The deformation terms of the energy line density in Eqs.~(\ref{Ebend}) and (\ref{Edisplacement}) are of quadratic order in the deflection $y(x)$. When one wishes to describe finite amplitudes, a nonlinearity is necessary to bound the deflection $y(x)$. Here, in the bulk terms, such a nonlinearity is only provided from the decreasing influence of the deflecting magnetic field in Eq.~(\ref{Emagnetic}). 
%In the present formulation, there are no nonlinear elastic bulk terms bounding the amplitude of deformation. Whether such nonlinear elastic effects play a role needs to be further investigated in the future.

Detailed knowledge about the boundary conditions of the deflection and its derivatives
 at the end points of the chain would be necessary to fully determine the chain shape
 from the above equations.
These boundary conditions depend on the interaction with the matrix and are not accessible in the present reduced framework.
Therefore, we proceed in a different way.
From the experiments, the shape of the chains is known and can to good approximation be represented
 by a polynomial form 
\begin{equation}\label{eqshape}
y(x) = S\prod_{m=0}^{M-1}(x-mb)\qquad\mbox{for }x_1\leq x\leq x_2,
\end{equation}
where $M$ is again the number of half-waves, the prefactor $S$ sets the strength or amount of chain deformation and deflection,
 $b$ is the spacing between the nodes, and the interval $[x_1,x_2]$ %is given by Eq.~(\ref{eqL}).
follows from the experimental result of preserved chain length $L$, 
\begin{equation}\label{eqL}
\int_{x_1}^{x_2}\sqrt{1+[y'(x)]^2} \:dx = L. 
\end{equation}
The polynomial form reproduces the experimentally observed straight chain ends as well as the smaller oscillation amplitudes inside longer chains.

Next, we insert Eq.~(\ref{eqshape}) into the above expressions for the energy and minimize with respect
 to $S$, $x_1$, and $x_2$ for a given $M$, with the constraint of constant length $L$, see Eq.~(\ref{eqL}).
Parameter values of the coefficients $C_b$ and $C_d$ %and $C_c$ 
are found by matching the resulting shapes
 to the corresponding experimental profiles (chain deformations for $G'=0.25\ \mathrm{Pa}$
 and magnetic field $B=100.8\ \mathrm{mT}$ as in Fig.~\ref{Fig.2}a, $M=2$, are used for this purpose).
We obtain \amm{$C_b\approx 0.01 Wb^2$ and $C_d\approx 2 W/b^2$.} %and $C_c\approx 1.5 W$.

\begin{figure}[t]
\centering
\includegraphics[width=8.6cm]{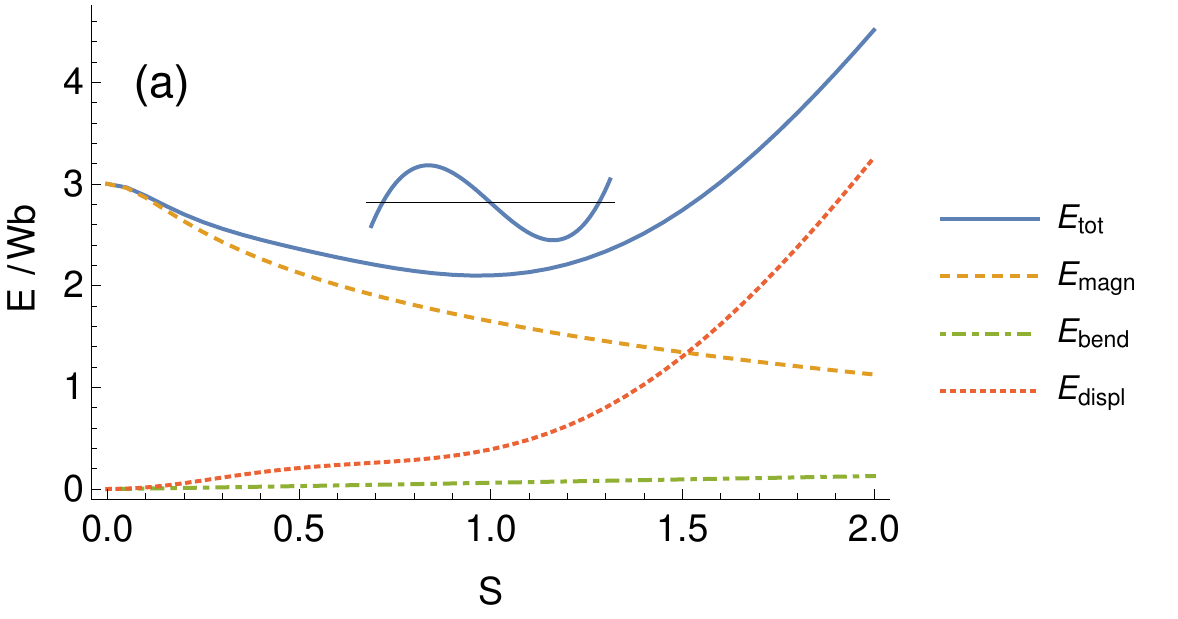}
\includegraphics[width=8.6cm]{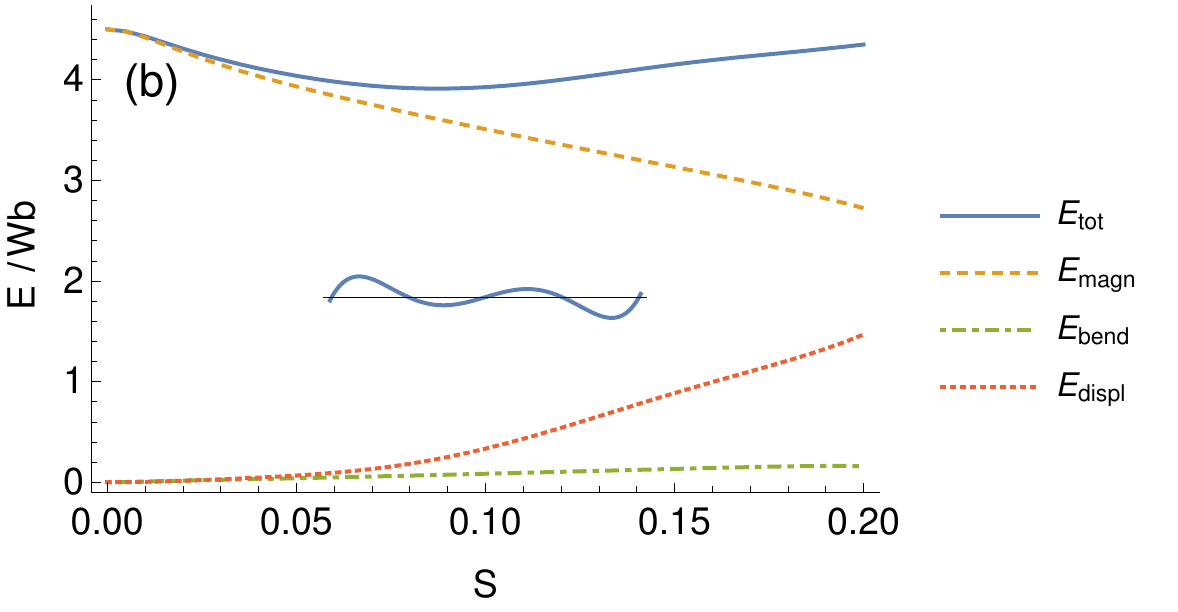}
\caption{Contributions to the total energy as a function of the strength $S$ of deformation and minimized
 with respect to $x_1$ and $x_2$ for a chain of the shape given by Eq.~(\ref{eqshape}).
Here we show the cases (a) $M=2$, $L=3b$ and (b) $M=4$, $L=4.5b$.
The total energy $E_{tot}$ has a global minimum as a function of $S$, which corresponds
 to the most stable chain shape.
%The ``bumps'' in the energetic contributions for small $S$ is attributed to a change in the kind of deformation. At small preset $S$, an asymmetric deformation is energetically preferred (left insets), whereas for larger $S$ this applies for symmetric shapes  (right insets).
We always observed the global minimum for symmetric shapes.}
\label{enva}
\end{figure}

To illustrate how the energetic contributions vary under increasing preset deformation,
 we plot in Fig.~\ref{enva} the energies for increasing $S$ for two fixed combinations of $M$ and $L$.
The total energy $E_{tot}$ shows a global minimum in both panels, which we always observed for symmetric chain deformations. 
As expected, with increasing amplitude $S$ the magnetic energy decreases, whereas the deformation
 energies increase.

\begin{figure}
\centering
\includegraphics[width=8.6cm]{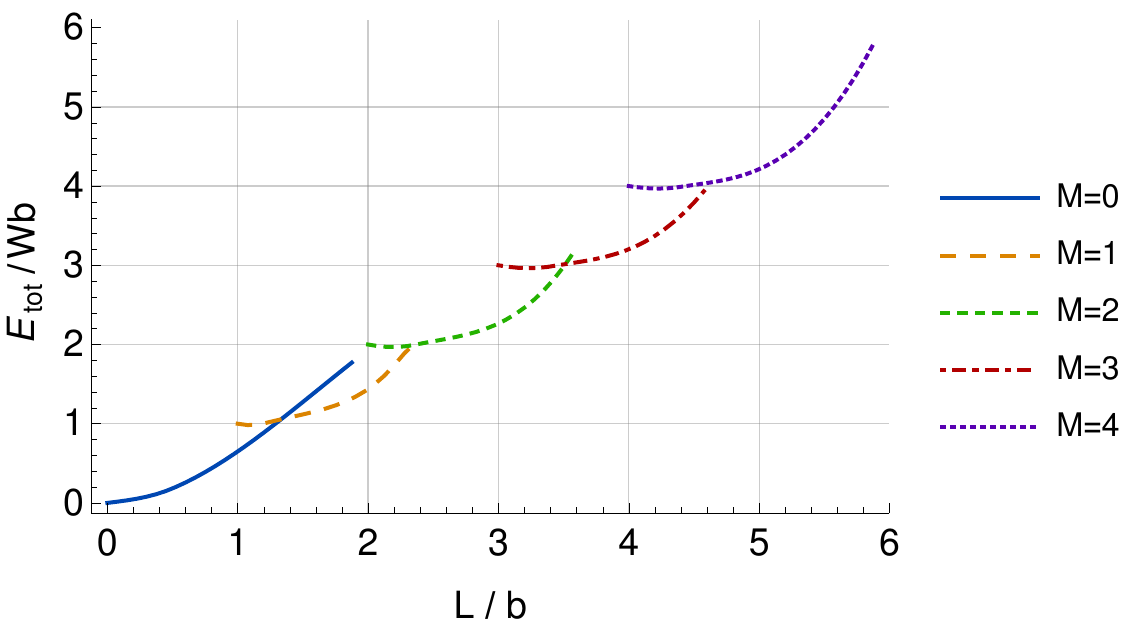}
\caption{Energies $E_{tot}$ of chain deformations of the shape given by Eq.~(\ref{eqshape}), minimized with respect
 to $S$, $x_1$, and $x_2$, as a function of chain length $L$ and number of half-oscillations $M$.
Each curve describes a shape of $M$ half-waves with a minimum total length of $(M-1)b$.
The resulting curves show crossing points from where the total energy for an increasing $L$ is lowered by bending one extra time (jumping to a higher $M$) rather than conserving the same shape.}
\label{energies}
\end{figure}

Next, we determine the minimal total energy as a function of chain length $L$ for increasing
 number of half-oscillations $M$.
We can see in Fig.~\ref{energies} that with increasing chain length $L$ the shapes that minimize the energy
 show an increasing number of half-waves $M$ in good agreement with the experimental data in Fig.~\ref{Fig.2}b.

\begin{figure}
\centering
\includegraphics[width=8.6cm]{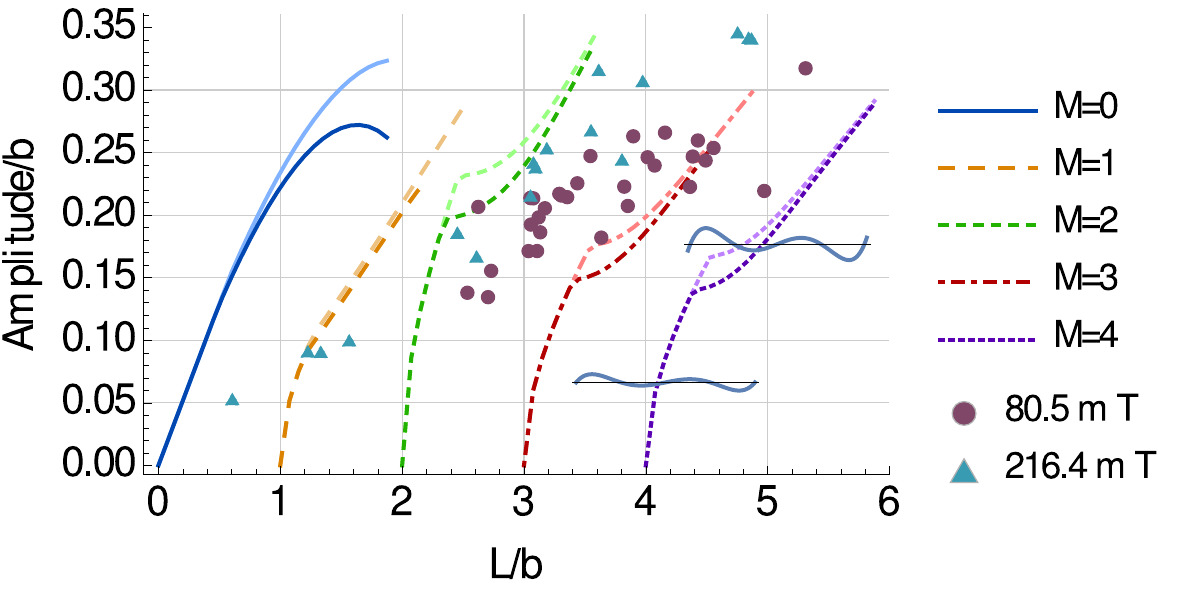}
\caption{Resulting deflection amplitudes of the chain deformation, calculated according to Eq.~(\ref{eqampl}).
Darker curves represent the model parameters optimized with respect to
 the experimental shapes for a magnetic field intensity $B=100.8\ \mathrm{mT}$.
Brighter curves were obtained by increasing the magnetic energy scale $W$ by a factor $\sim 1.42$,
 which corresponds %, according to the magnetization depicted in Fig.~S1b,
 to an increased magnetic
 field intensity of $B\sim 216\ \mathrm{mT}$ \cite{suppl}, comparable with the triangular experimental data points.
Both, model curves and included experimental data points, demonstrate that moderate variations of the magnetic field intensities only slightly affect the observed deflection and deformation amplitudes. 
The value of $b$ necessary to perform the analysis was determined from the $M=2$ experimental data as $b=12.6\ \mathrm{\mu m}$.  
For $M\geq2$ ``bumps'' appear in the curves, which arise from a change in the type of deformation as illustrated
 by the insets: for each $M$ curve, the end points of the chain for lower $L$ correspond to nodes of the deformational oscillations, i.e.\ 
 $y(x_1)\simeq y(x_2)\simeq 0$ (lower left inset); for higher $L$, these outer nodes shift to the inside of the chain (upper right inset).
As seen from Fig.~\ref{energies}, the lower-$L$ parts of the curves are not energetically preferred.}
\label{amplitudes}
\end{figure}

Moreover, we quantify the amplitude of the chain deflection or deformation by
\begin{equation}\label{eqampl}
\mathrm{Amplitude}=\sqrt{  \langle y^2 \rangle - {\langle y \rangle}^2 }, \qquad
 \langle\cdot\rangle=\frac{\int_{x_1}^{x_2}{\cdot\ dx}}{x_2-x_1}.
\end{equation}
Resulting values are plotted in Fig.~\ref{amplitudes} and compared with corresponding experimental data.
As mentioned above, we optimized the model parameters with respect to the experimental data for a magnetic field intensity of $B=100.8\ \mathrm{mT}$. 
We demonstrate in Fig.~\ref{amplitudes} that moderate variations of the magnetic field intensity only slightly affect our results: the brighter curves are obtained when multiplying the magnetic energy scale $W$ by a factor $\sim 1.42$, corresponding to an increased magnetic field intensity of approximately $B\sim 216\ \mathrm{mT}$ \cite{suppl}. This is in agreement with the experimental observations. We include in Fig.~\ref{amplitudes} the experimentally determined values for $B= 80.5\ \mathrm{mT}$ and $B= 216.4\ \mathrm{mT}$. Only a slight trend of increasing deflection amplitudes is found for this increase of magnetic field intensity. 

% intermediate between the two experimental data sets shown in Fig.~\ref{amplitudes}. Therefore, to compare with the data corresponding to ${B}=216.4\ \mathrm{mT}$ we multiply $W$ by a factor $\sim 1.4$. To make a proper comparison, we set $b=12.6 \mathrm{\mu m}$, which is the spacing extrapolated for the $M=2$ experimental shapes. Nevertheless, the $b$ parameter can depend on $M$, and indeed the experiments suggest that it tends to increase with the number of half-oscillations. In the region between $L\sim 2.5b$ and $L\sim 4.5b$ the model curves show a good agreement with the data. Moreover, theoretically and experimentally we observe the amplitude of the deformation to increase with increasing magnetic interaction. An upper limitation of the feasible amplitude is also provided by this model.
Together, although the curves for $M=2$ in Fig.~\ref{amplitudes} slightly overshoot the data points, Figs.~\ref{energies} and \ref{amplitudes} are in good agreement with the experimental results. 
The amplitude of deflection and deformation is not observed to unboundedly increase with chain length $L$ in the experiments. 
Likewise, our model predicts that longer chains prefer to bend one extra time (switching
 to higher-$M$ shape) rather than to show too large deflection amplitudes.

\section{Coarse-grained molecular dynamics simulation}

We also studied the buckling of the chain using two-dimensional coarse-grained molecular dynamics simulations by means of the ESPResSo software \cite{limbach06a,arnold13a}.
A simple model was developed that allowed us to analyze the influence of particular interactions and material properties on the buckling effect. Here, we focus on the elasticity of the polymer matrix in the immediate vicinity of the magnetic particles.

By choosing the coarse-grained scale for our model, we ignore any chemical details but rather describe the system in terms of the magnetic particles as well as small pieces of polymer gel. As the buckling effect appears to be two-dimensional, and as the ground states for systems of dipolar particles have also been found to be two-dimensional \cite{prokopieva09a}, we use this dimensionality for our simulations. We study a chain of 100 magnetic particles with 
a significant amount of surrounding elastic matrix. 
%enough surrounding gel to allow the strain field to decay to XX\% at the boundary. \gka{Rudolf could you give a number here. }

As in the analytical approach, the gel matrix is modeled by a network of springs. Here, however, we use a regular hexagonal mesh as a basis. To mimic the non-linear elastic behavior of polymers, we use a finitely extensible non-linear elastic spring potential (FENE-potential \cite{warner1972kinetic}) for the springs along the edges of the mesh. As a simple implementation of the finite compressibility, we introduce FENE-like angular potentials with a divergence at zero and 180 degrees on the angles at the mesh points \cite{suppl}. The magnetic particles are modeled as rigid spheres interacting by a truncated, purely repulsive Lennard-Jones potential, the so-called Weeks-Chandler-Andersen potential \cite{weeks71a,suppl}. %\gka{\em Two Question: How do we reference to the supplementary material? Should mention that the absolute numbers of the gel modulus might not be the same in the simulation and the experiments ? }
%???suppl
Their magnetic moment is assumed to be determined purely by the external magnetic field and to be constant throughout the simulation. I.e.\ we assume that the external field is significantly stronger than the field created by the particles.
The magnetic moments are taken parallel to the external field and with a magnitude given by the experimentally observed magnetization curve.
%The magnetization can rotate freely in the particles. \gka{\em Question: Rudolf is this correct ? }
The coupling between the particles and the mesh is introduced in such a way, that under the volume occupied by a particle, the mesh does not deform, but rigidly follows the translational and rotational motion of the particle \cite{suppl}. Hence, a local shear strain on the matrix can result in the rotation of a magnetic particle, but not its magnetic moment.
%???supplementary information

An important point is the elasticity of the polymer matrix in the immediate vicinity of the magnetic particles and, in particular, between two magnetic particles. 
We study two situations here, the first one including a stiffer region in the immediate vicinity of the particles, the second one without such a stiffer layer and directly jumping to the bulk elasticity.
The stiffer layer, if imposed, is created using a spring constant larger by four orders of magnitude on those springs which originate from mesh sites within the particle volumes \cite{suppl}. The angular potentials are unchanged.
%???cite suppl

\begin{figure}
\begin{center}
\includegraphics[width=0.8\linewidth]{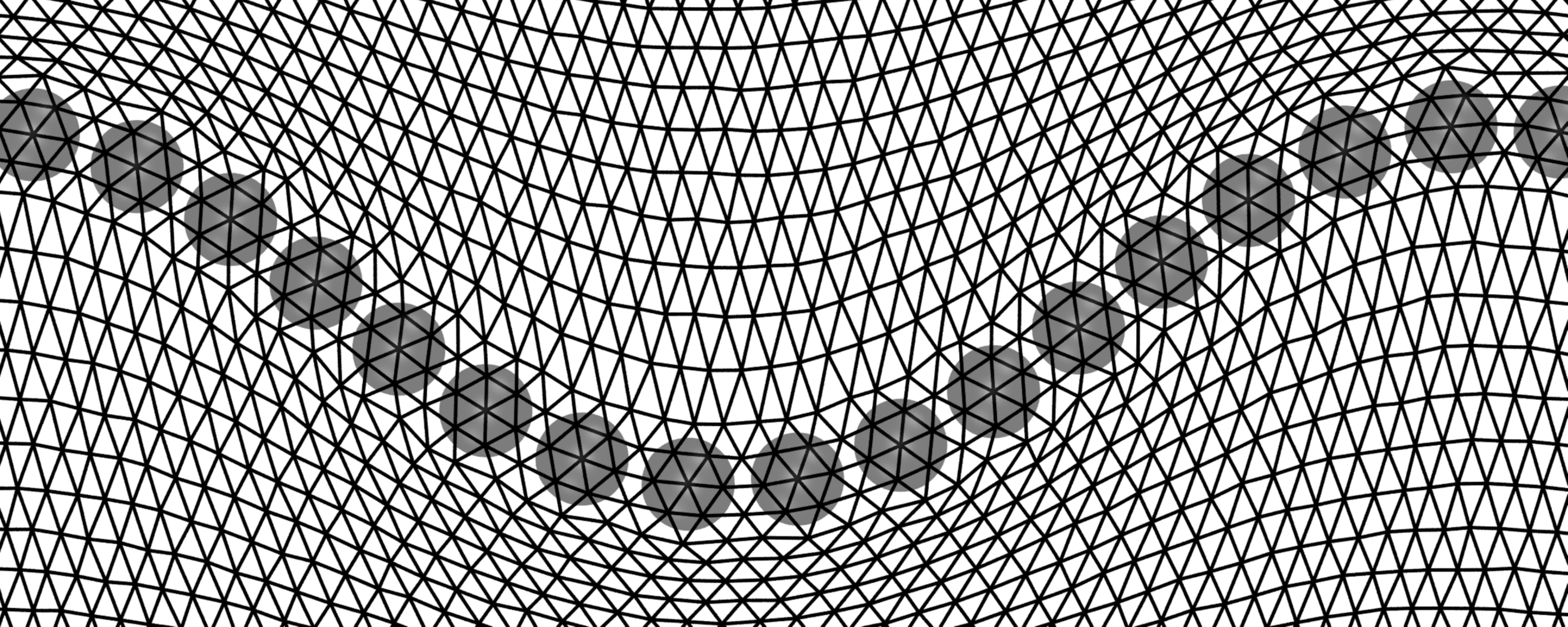}
\\[.1cm]
\includegraphics[width=0.8\linewidth]{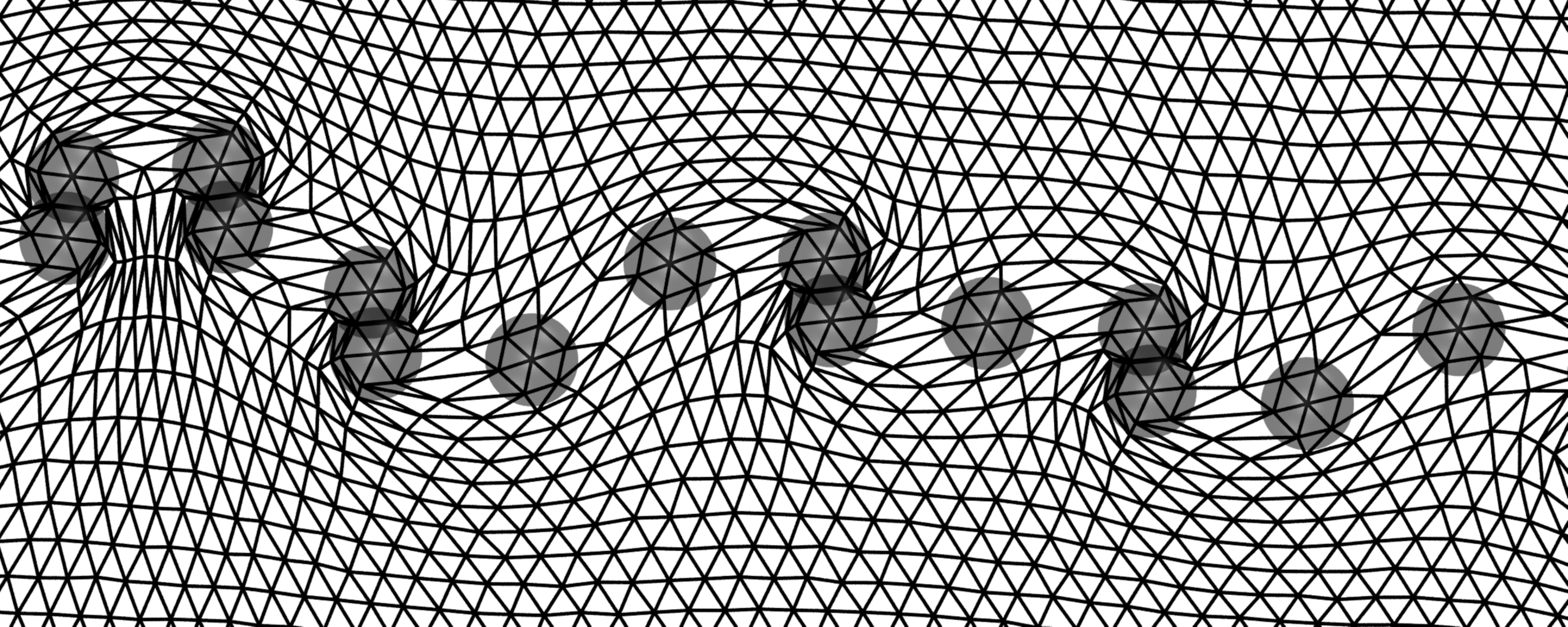}
\end{center}
\caption{
\label{fig:boundary-layer}
Detailed view of the local deformations in the polymer mesh around the magnetic particles with a layer of increased stiffness (top) and without one (bottom) in the immediate vicinity of the particle surfaces. The external magnetic field of strength $216\;\mathrm{mT}$ is applied in the vertical direction.
When the boundary layer is assumed to be stiffer than the bulk (top), the buckling effect, as observed in the experiments, occurs.
When the layer around the particles is soft (bottom), neighboring particles either form tight columns parallel to the field, or separate in the direction perpendicular to the field.
Both, the spacing between neighboring particles in the case of a stiff surface layer (top) and the overlap of the rigid magnetic particles in the case of a soft surface layer (bottom) can be explained by a very strong dipole-dipole interaction, which is repulsive in the one case and attractive in the other.
}
\end{figure}

\begin{figure}
\begin{center}
%\includegraphics[width=\linewidth]{w-stiffness-50mt.pdf}
%\\
\vspace{0.3em}
\includegraphics[width=\linewidth]{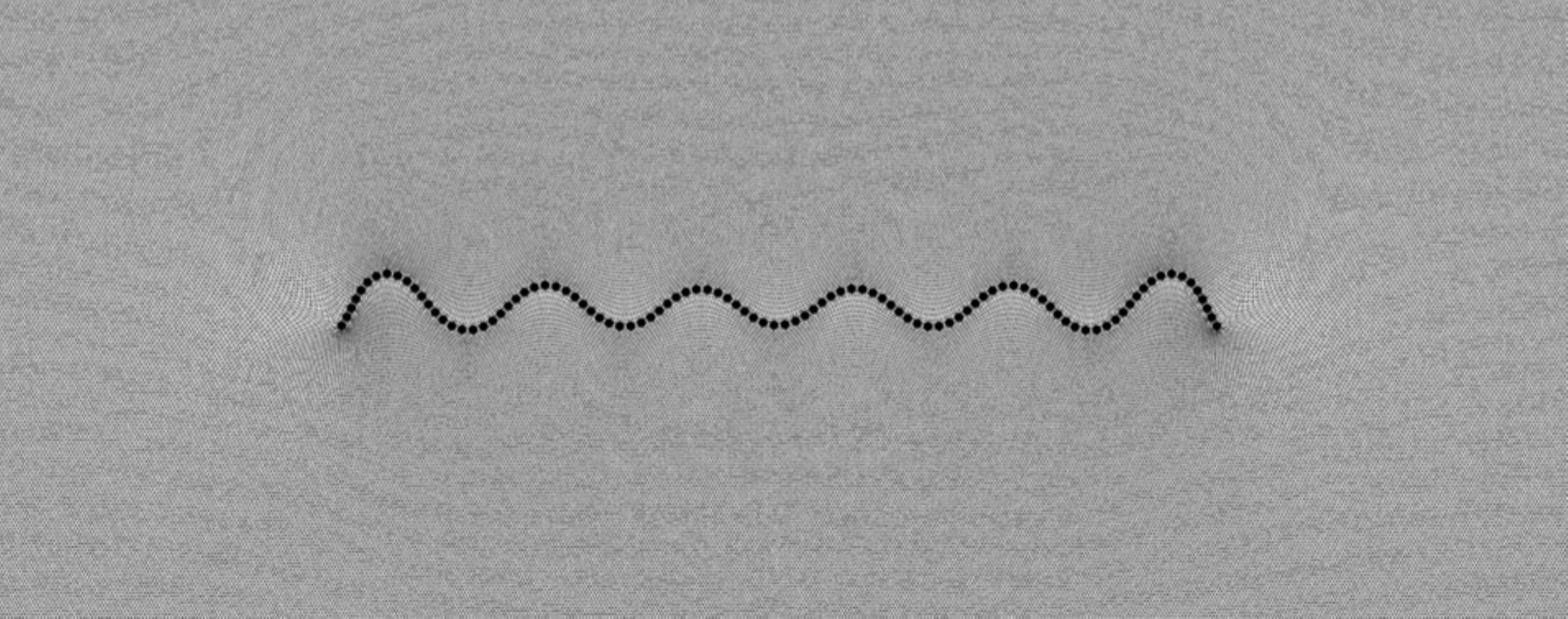}
\end{center}
\caption{\label{fig:sim-buckling-chains}
Buckling chain of magnetic particles and the surrounding polymer mesh for an external field of magnitude $216\;\mathrm{mT}$ pointing along the vertical direction.
In this image, roughly one quarter of the full simulation area is shown. 
%The amplitude of the chain oscillation increases with increasing induced dipole moments of the magnetic particles. 
The surrounding matrix follows the chain oscillation with an amplitude that decreases over distance from the chain.
%\gka{\em Remark: In the main text I would only show one of the figures. We can however include both figures in the supplementary material. }
}
\end{figure}

A comparison between the cases with and without a stiffer layer of gel around the magnetic particles can be seen in Fig.~\ref{fig:boundary-layer}.
The images show a small part of the resulting configuration of magnetic particles and the surrounding mesh for a field applied perpendicular to the initial chain direction.
Thus the magnetic moments of the particles are oriented perpendicular to the undistorted chain direction. This results in an energetically unfavorable parallel side-by-side configuration for the dipole moments. The energy can be reduced either by increasing the distance between the dipoles along the initial chain direction, or by moving dipoles perpendicularly to the initial chain direction
so that they approach the energetically most favorable head-to-tail configuration. 
If the matrix is made stiffer immediately around the particles, and thus the contour length of the chain cannot change significantly, the re-positioning towards the head-to-tail configuration causes the buckling effect seen in the experiments (Fig.~\ref{fig:boundary-layer}).
When one assumes the matrix immediately around the magnetic particles to be as soft as in the bulk of the material, neighboring particles can move apart and the chain breaks up into individual particles or small columns perpendicular to the original chain direction. Additionally, a layer of increased stiffness also introduces a bending rigidity of the chain. 
In Fig.~\ref{fig:sim-buckling-chains}, the full chain and the surrounding matrix is shown for an external field of magnitude $216\ \mathrm{mT}$, which from the experimental magnetization measurements corresponds to a magnetic moment of about $4.5 \cdot 10^{-14}\ \mathrm{Am}^2$ \cite{suppl}. 
Due to the different dimensionalities, the elastic modulus of the surrounding matrix could not be directly matched to the experimental system. 
%\gka{\em Remark: Since we don't know the elastic modulus, I would not give the magnetic moment. }

Actually, the amplitude of the chain oscillation increases when the external field is higher and induces larger dipole moments in the particles. This increases the tendency of the magnetic moments to approach the head-to-tail configuration, which in turn leads to a stronger deformation of the matrix. We note that the relative amplitude of the buckling along the chain is similar in the simulations (Fig.~\ref{fig:boundary-layer}) and experiments (Fig.~\ref{Fig.2}).
The matrix surrounding the chain follows the chain oscillation with an amplitude that decreases over distance from the chain. 
Deviations may be expected from the deformational
far-field in the experimental system due to the different dimensionalities of the systems.


\section{Conclusions}
We have shown that paramagnetic chains in a soft polymer gel can buckle in a perpendicular magnetic field. The buckling morphology depends on the length of the chain, the strength of the magnetic field and the modulus of the gel. Longer chains form buckling structures with a higher number of half waves. \amm{Higher strengths} of the magnetic field and a lower modulus of the gel matrix \amm{can} lead to \amm{higher deformation amplitudes}. The deformation field in the \amm{surrounding gel matrix} confirms that the \amm{embedding} polymer network is strongly coupled to the paramagnetic chain. A \amm{minimal} magneto-elastic coupling model is developed to describe the \amm{morphological} behavior of the paramagnetic chains in \amm{the} soft gel under a perpendicular magnetic field. It shows that the chains deform in order to decrease the magnetic energy. \amm{This is hindered by the simultaneous deformation of the gel matrix, which increases the elastic energy of the gel.} 
%We also find that the paramagnetic chains have a bending rigidity on the order of $10^{-21}-10^{-20}$  Nm$^2$. This bending rigidity impedes the deformation of the paramagnetic chains. 
\gka{
Additionally, we have introduced a coarse-grained molecular dynamics simulation model, which covers both, the magnetic particles and the surrounding polymer mesh. In this model,  the buckling of the chains can only be observed when the surface layer around the particles is assumed to be stiffer than the bulk of the gel. This prevents the chain from breaking up into columns oriented perpendicular to the initial chain direction or into isolated particles.} %In addition, it creates a bending rigidity which we also use in the analytical model.
\amm{These findings support the picture that the embedded magnetic chains themselves feature a certain bending rigidity, possibly due to polymer chains adsorbed on the particle surfaces.} 

\amm{Since} the magneto-elastic effect \amm{demonstrated and analyzed} in this paper is pronounced, reversible and controllable, it may be useful for designing micro-devices, e.g. micro-valves and pumps for microfluidic control \amm{\cite{fleischmann2012one}}. As the morphologies of the buckling paramagnetic chains are correlated with the modulus of the gel matrix, we may use them as mechanical probes for soft gels (similarly to active microrheology techniques) \amm{\cite{Wilhelm2008}}. Moreover, our study may help to understand the physical interactions between the magnetic chains and the surrounding cytoskeleton network in magnetotactic bacteria \amm{\cite{Kornig2014}}. In our future study we will focus on how the interfacial coupling between the magnetic particles and the polymer network influences the local magneto-elastic coupling effect.

\section{Acknowledgements}
\amm{The authors} thank Dr.~Peter \amm{Bl\"umler} for inspiring discussions \amm{on} designing the Halbach magnetic arrays \amm{and} acknowledge \amm{support} by the Deutsche Forschungsgemeinschaft (DFG) through the SPP 1681.
\rw{RW and CH acknowledge funding through the cluster of excellence EXC 310, SimTech at the University of Stuttgart.}

%\bibliography{lit_rsc} %your .bib file

\begin{thebibliography}{62}
\expandafter\ifx\csname natexlab\endcsname\relax\def\natexlab#1{#1}\fi
\expandafter\ifx\csname bibnamefont\endcsname\relax
  \def\bibnamefont#1{#1}\fi
\expandafter\ifx\csname bibfnamefont\endcsname\relax
  \def\bibfnamefont#1{#1}\fi
\expandafter\ifx\csname citenamefont\endcsname\relax
  \def\citenamefont#1{#1}\fi
\expandafter\ifx\csname url\endcsname\relax
  \def\url#1{\texttt{#1}}\fi
\expandafter\ifx\csname urlprefix\endcsname\relax\def\urlprefix{URL }\fi
\providecommand{\bibinfo}[2]{#2}
\providecommand{\eprint}[2][]{\url{#2}}

\bibitem[{\citenamefont{Ilg}(2013)}]{Ilg2013}
\bibinfo{author}{\bibfnamefont{P.}~\bibnamefont{Ilg}}, \bibinfo{journal}{Soft
  Matter} \textbf{\bibinfo{volume}{9}}, \bibinfo{pages}{3465}
  (\bibinfo{year}{2013}).

\bibitem[{\citenamefont{Snyder et~al.}(2010)\citenamefont{Snyder, Nguyen, and
  Ramanujan}}]{Snyder2010}
\bibinfo{author}{\bibfnamefont{R.~L.} \bibnamefont{Snyder}},
  \bibinfo{author}{\bibfnamefont{V.~Q.} \bibnamefont{Nguyen}},
  \bibnamefont{and} \bibinfo{author}{\bibfnamefont{R.~V.}
  \bibnamefont{Ramanujan}}, \bibinfo{journal}{Smart Mater. Struct.}
  \textbf{\bibinfo{volume}{19}}, \bibinfo{pages}{055017}
  (\bibinfo{year}{2010}).

\bibitem[{\citenamefont{Zimmermann et~al.}(2006)\citenamefont{Zimmermann,
  Naletova, Zeidis, B{\"o}hm, and Kolev}}]{zimmermann2006modelling}
\bibinfo{author}{\bibfnamefont{K.}~\bibnamefont{Zimmermann}},
  \bibinfo{author}{\bibfnamefont{V.~A.} \bibnamefont{Naletova}},
  \bibinfo{author}{\bibfnamefont{I.}~\bibnamefont{Zeidis}},
  \bibinfo{author}{\bibfnamefont{V.}~\bibnamefont{B{\"o}hm}}, \bibnamefont{and}
  \bibinfo{author}{\bibfnamefont{E.}~\bibnamefont{Kolev}}, \bibinfo{journal}{J.
  Phys.: Condens. Matter} \textbf{\bibinfo{volume}{18}}, \bibinfo{pages}{S2973}
  (\bibinfo{year}{2006}).

\bibitem[{\citenamefont{Hergt et~al.}(2006)\citenamefont{Hergt, Dutz,
  M{\"u}ller, and Zeisberger}}]{hergt2006magnetic}
\bibinfo{author}{\bibfnamefont{R.}~\bibnamefont{Hergt}},
  \bibinfo{author}{\bibfnamefont{S.}~\bibnamefont{Dutz}},
  \bibinfo{author}{\bibfnamefont{R.}~\bibnamefont{M{\"u}ller}},
  \bibnamefont{and}
  \bibinfo{author}{\bibfnamefont{M.}~\bibnamefont{Zeisberger}},
  \bibinfo{journal}{J. Phys.: Condens. Matter} \textbf{\bibinfo{volume}{18}},
  \bibinfo{pages}{S2919} (\bibinfo{year}{2006}).

\bibitem[{\citenamefont{Szab\'o et~al.}(1998)\citenamefont{Szab\'o, Szeghy, and
  Zr\'inyi}}]{Szab1998}
\bibinfo{author}{\bibfnamefont{D.}~\bibnamefont{Szab\'o}},
  \bibinfo{author}{\bibfnamefont{G.}~\bibnamefont{Szeghy}}, \bibnamefont{and}
  \bibinfo{author}{\bibfnamefont{M.}~\bibnamefont{Zr\'inyi}},
  \bibinfo{journal}{Macromolecules} \textbf{\bibinfo{volume}{31}},
  \bibinfo{pages}{6541} (\bibinfo{year}{1998}).

\bibitem[{\citenamefont{Abramchuk et~al.}(2006)\citenamefont{Abramchuk,
  Grishin, Kramarenko, Stepanov, and Khokhlov}}]{Abramchuk2006}
\bibinfo{author}{\bibfnamefont{S.~S.} \bibnamefont{Abramchuk}},
  \bibinfo{author}{\bibfnamefont{D.~A.} \bibnamefont{Grishin}},
  \bibinfo{author}{\bibfnamefont{E.~Y.} \bibnamefont{Kramarenko}},
  \bibinfo{author}{\bibfnamefont{G.~V.} \bibnamefont{Stepanov}},
  \bibnamefont{and} \bibinfo{author}{\bibfnamefont{A.~R.}
  \bibnamefont{Khokhlov}}, \bibinfo{journal}{Polym. Sci. Ser. A}
  \textbf{\bibinfo{volume}{48}}, \bibinfo{pages}{138} (\bibinfo{year}{2006}).

\bibitem[{\citenamefont{Filipcsei
  et~al.}(2007{\natexlab{a}})\citenamefont{Filipcsei, Csetneki, Szil{\'a}gyi,
  and Zr\'{i}nyi}}]{filipcsei2007magnetic}
\bibinfo{author}{\bibfnamefont{G.}~\bibnamefont{Filipcsei}},
  \bibinfo{author}{\bibfnamefont{I.}~\bibnamefont{Csetneki}},
  \bibinfo{author}{\bibfnamefont{A.}~\bibnamefont{Szil{\'a}gyi}},
  \bibnamefont{and}
  \bibinfo{author}{\bibfnamefont{M.}~\bibnamefont{Zr\'{i}nyi}},
  \bibinfo{journal}{Adv. Polym. Sci.} \textbf{\bibinfo{volume}{206}},
  \bibinfo{pages}{137} (\bibinfo{year}{2007}{\natexlab{a}}).

\bibitem[{\citenamefont{Collin et~al.}(2003)\citenamefont{Collin, Auernhammer,
  Gavat, Martinoty, and Brand}}]{Collin2003}
\bibinfo{author}{\bibfnamefont{D.}~\bibnamefont{Collin}},
  \bibinfo{author}{\bibfnamefont{G.~K.} \bibnamefont{Auernhammer}},
  \bibinfo{author}{\bibfnamefont{O.}~\bibnamefont{Gavat}},
  \bibinfo{author}{\bibfnamefont{P.}~\bibnamefont{Martinoty}},
  \bibnamefont{and} \bibinfo{author}{\bibfnamefont{H.~R.} \bibnamefont{Brand}},
  \bibinfo{journal}{Macromol. Rapid Comm.} \textbf{\bibinfo{volume}{24}},
  \bibinfo{pages}{737} (\bibinfo{year}{2003}).

\bibitem[{\citenamefont{Faraudo et~al.}(2013)\citenamefont{Faraudo, Andreu, and
  Camacho}}]{Faraudo2013}
\bibinfo{author}{\bibfnamefont{J.}~\bibnamefont{Faraudo}},
  \bibinfo{author}{\bibfnamefont{J.~S.} \bibnamefont{Andreu}},
  \bibnamefont{and} \bibinfo{author}{\bibfnamefont{J.}~\bibnamefont{Camacho}},
  \bibinfo{journal}{Soft Matter} \textbf{\bibinfo{volume}{9}},
  \bibinfo{pages}{6654} (\bibinfo{year}{2013}).

\bibitem[{\citenamefont{Griffiths}(1999)}]{Griffiths1999}
\bibinfo{author}{\bibfnamefont{D.~J.} \bibnamefont{Griffiths}},
  \emph{\bibinfo{title}{Introduction to Electrodynamics}}
  (\bibinfo{publisher}{Prentice-Hall, Upper Saddle River, NJ,},
  \bibinfo{year}{1999}), \bibinfo{edition}{3rd} ed.

\bibitem[{\citenamefont{Klapp and Schoen}(2002)}]{klapp2002spontaneous}
\bibinfo{author}{\bibfnamefont{S.~H.~L.} \bibnamefont{Klapp}} \bibnamefont{and}
  \bibinfo{author}{\bibfnamefont{M.}~\bibnamefont{Schoen}},
  \bibinfo{journal}{J. Chem. Phys.} \textbf{\bibinfo{volume}{117}},
  \bibinfo{pages}{8050} (\bibinfo{year}{2002}).

\bibitem[{\citenamefont{Klapp}(2005)}]{klapp2005dipolar}
\bibinfo{author}{\bibfnamefont{S.~H.~L.} \bibnamefont{Klapp}},
  \bibinfo{journal}{J. Phys.: Condens. Matter} \textbf{\bibinfo{volume}{17}},
  \bibinfo{pages}{R525} (\bibinfo{year}{2005}).

\bibitem[{\citenamefont{Gajula et~al.}(2010)\citenamefont{Gajula,
  Neves-Petersen, and Petersen}}]{Gajula2010}
\bibinfo{author}{\bibfnamefont{G.~P.} \bibnamefont{Gajula}},
  \bibinfo{author}{\bibfnamefont{M.~T.} \bibnamefont{Neves-Petersen}},
  \bibnamefont{and} \bibinfo{author}{\bibfnamefont{S.~B.}
  \bibnamefont{Petersen}}, \bibinfo{journal}{Appl. Phys. Lett.}
  \textbf{\bibinfo{volume}{97}}, \bibinfo{pages}{103103}
  (\bibinfo{year}{2010}).

\bibitem[{\citenamefont{de~Vicente et~al.}(2011)\citenamefont{de~Vicente,
  Klingenberg, and Hidalgo-Alvarez}}]{deVicente2011}
\bibinfo{author}{\bibfnamefont{J.}~\bibnamefont{de~Vicente}},
  \bibinfo{author}{\bibfnamefont{D.~J.} \bibnamefont{Klingenberg}},
  \bibnamefont{and}
  \bibinfo{author}{\bibfnamefont{R.}~\bibnamefont{Hidalgo-Alvarez}},
  \bibinfo{journal}{Soft Matter} \textbf{\bibinfo{volume}{7}},
  \bibinfo{pages}{3701} (\bibinfo{year}{2011}).

\bibitem[{\citenamefont{Auernhammer et~al.}(2006)\citenamefont{Auernhammer,
  Collin, and Martinoty}}]{Auernhammer2006}
\bibinfo{author}{\bibfnamefont{G.~K.} \bibnamefont{Auernhammer}},
  \bibinfo{author}{\bibfnamefont{D.}~\bibnamefont{Collin}}, \bibnamefont{and}
  \bibinfo{author}{\bibfnamefont{P.}~\bibnamefont{Martinoty}},
  \bibinfo{journal}{J. Chem. Phys.} \textbf{\bibinfo{volume}{124}},
  \bibinfo{pages}{204907} (\bibinfo{year}{2006}).

\bibitem[{\citenamefont{Wood and Camp}(2011)}]{wood2011modeling}
\bibinfo{author}{\bibfnamefont{D.~S.} \bibnamefont{Wood}} \bibnamefont{and}
  \bibinfo{author}{\bibfnamefont{P.~J.} \bibnamefont{Camp}},
  \bibinfo{journal}{Phys. Rev. E} \textbf{\bibinfo{volume}{83}},
  \bibinfo{pages}{011402} (\bibinfo{year}{2011}).

\bibitem[{\citenamefont{Ivaneyko et~al.}(2012)\citenamefont{Ivaneyko,
  Toshchevikov, Saphiannikova, and Heinrich}}]{ivaneyko2012effects}
\bibinfo{author}{\bibfnamefont{D.}~\bibnamefont{Ivaneyko}},
  \bibinfo{author}{\bibfnamefont{V.}~\bibnamefont{Toshchevikov}},
  \bibinfo{author}{\bibfnamefont{M.}~\bibnamefont{Saphiannikova}},
  \bibnamefont{and} \bibinfo{author}{\bibfnamefont{G.}~\bibnamefont{Heinrich}},
  \bibinfo{journal}{Condens. Matter Phys.} \textbf{\bibinfo{volume}{15}},
  \bibinfo{pages}{33601} (\bibinfo{year}{2012}).

\bibitem[{\citenamefont{Han et~al.}(2013)\citenamefont{Han, Hong, and
  Faidley}}]{Han2013}
\bibinfo{author}{\bibfnamefont{Y.}~\bibnamefont{Han}},
  \bibinfo{author}{\bibfnamefont{W.}~\bibnamefont{Hong}}, \bibnamefont{and}
  \bibinfo{author}{\bibfnamefont{L.~E.} \bibnamefont{Faidley}},
  \bibinfo{journal}{Int. J. Solids Struct.} \textbf{\bibinfo{volume}{50}},
  \bibinfo{pages}{2281} (\bibinfo{year}{2013}).

\bibitem[{\citenamefont{Pessot et~al.}(2014)\citenamefont{Pessot, Cremer,
  Borin, Odenbach, L\"owen, and Menzel}}]{Pessot2014}
\bibinfo{author}{\bibfnamefont{G.}~\bibnamefont{Pessot}},
  \bibinfo{author}{\bibfnamefont{P.}~\bibnamefont{Cremer}},
  \bibinfo{author}{\bibfnamefont{D.~Y.} \bibnamefont{Borin}},
  \bibinfo{author}{\bibfnamefont{S.}~\bibnamefont{Odenbach}},
  \bibinfo{author}{\bibfnamefont{H.}~\bibnamefont{L\"owen}}, \bibnamefont{and}
  \bibinfo{author}{\bibfnamefont{A.~M.} \bibnamefont{Menzel}},
  \bibinfo{journal}{J. Chem. Phys.} \textbf{\bibinfo{volume}{141}},
  \bibinfo{pages}{124904} (\bibinfo{year}{2014}).

\bibitem[{\citenamefont{Menzel}(2015)}]{Menzel2015}
\bibinfo{author}{\bibfnamefont{A.~M.} \bibnamefont{Menzel}},
  \bibinfo{journal}{Phys. Rep.} \textbf{\bibinfo{volume}{554}},
  \bibinfo{pages}{1} (\bibinfo{year}{2015}).

\bibitem[{\citenamefont{Stolbov et~al.}(2011)\citenamefont{Stolbov, Raikher,
  and Balasoiu}}]{stolbov11a}
\bibinfo{author}{\bibfnamefont{O.~V.} \bibnamefont{Stolbov}},
  \bibinfo{author}{\bibfnamefont{Y.~L.} \bibnamefont{Raikher}},
  \bibnamefont{and} \bibinfo{author}{\bibfnamefont{M.}~\bibnamefont{Balasoiu}},
  \bibinfo{journal}{Soft Matter} \textbf{\bibinfo{volume}{7}},
  \bibinfo{pages}{8484} (\bibinfo{year}{2011}).

\bibitem[{\citenamefont{Filipcsei
  et~al.}(2007{\natexlab{b}})\citenamefont{Filipcsei, Csetneki, Szil\'agyi, and
  Zr\'inyi}}]{Filipcsei2007}
\bibinfo{author}{\bibfnamefont{G.}~\bibnamefont{Filipcsei}},
  \bibinfo{author}{\bibfnamefont{I.}~\bibnamefont{Csetneki}},
  \bibinfo{author}{\bibfnamefont{A.}~\bibnamefont{Szil\'agyi}},
  \bibnamefont{and} \bibinfo{author}{\bibfnamefont{M.}~\bibnamefont{Zr\'inyi}},
  \emph{\bibinfo{title}{Magnetic Field-Responsive Smart Polymer Composites}}
  (\bibinfo{publisher}{Springer Berlin Heidelberg},
  \bibinfo{year}{2007}{\natexlab{b}}), vol. \bibinfo{volume}{206} of
  \emph{\bibinfo{series}{Adv. Polym. Sci.}}, pp. \bibinfo{pages}{137--189}.

\bibitem[{\citenamefont{G\"unther et~al.}(2012)\citenamefont{G\"unther, Borin,
  G\"unther, and Odenbach}}]{Gunther2012}
\bibinfo{author}{\bibfnamefont{D.}~\bibnamefont{G\"unther}},
  \bibinfo{author}{\bibfnamefont{D.~Y.} \bibnamefont{Borin}},
  \bibinfo{author}{\bibfnamefont{S.}~\bibnamefont{G\"unther}},
  \bibnamefont{and} \bibinfo{author}{\bibfnamefont{S.}~\bibnamefont{Odenbach}},
  \bibinfo{journal}{Smart Mater. Struct.} \textbf{\bibinfo{volume}{21}},
  \bibinfo{pages}{015005} (\bibinfo{year}{2012}).

\bibitem[{\citenamefont{Borb{\'a}th et~al.}(2012)\citenamefont{Borb{\'a}th,
  G{\"u}nther, Borin, Gundermann, and Odenbach}}]{borbath2012xmuct}
\bibinfo{author}{\bibfnamefont{T.}~\bibnamefont{Borb{\'a}th}},
  \bibinfo{author}{\bibfnamefont{S.}~\bibnamefont{G{\"u}nther}},
  \bibinfo{author}{\bibfnamefont{D.~Y.} \bibnamefont{Borin}},
  \bibinfo{author}{\bibfnamefont{T.}~\bibnamefont{Gundermann}},
  \bibnamefont{and} \bibinfo{author}{\bibfnamefont{S.}~\bibnamefont{Odenbach}},
  \bibinfo{journal}{Smart Mater. Struct.} \textbf{\bibinfo{volume}{21}},
  \bibinfo{pages}{105018} (\bibinfo{year}{2012}).

\bibitem[{\citenamefont{Guan et~al.}(2008)\citenamefont{Guan, Dong, and
  Ou}}]{Guan2008}
\bibinfo{author}{\bibfnamefont{X.~C.} \bibnamefont{Guan}},
  \bibinfo{author}{\bibfnamefont{X.~F.} \bibnamefont{Dong}}, \bibnamefont{and}
  \bibinfo{author}{\bibfnamefont{J.~P.} \bibnamefont{Ou}}, \bibinfo{journal}{J.
  Magn. Magn. Mater.} \textbf{\bibinfo{volume}{320}}, \bibinfo{pages}{158}
  (\bibinfo{year}{2008}).

\bibitem[{\citenamefont{Danas et~al.}(2012)\citenamefont{Danas, Kankanala, and
  Triantafyllidis}}]{Danas2012}
\bibinfo{author}{\bibfnamefont{K.}~\bibnamefont{Danas}},
  \bibinfo{author}{\bibfnamefont{S.~V.} \bibnamefont{Kankanala}},
  \bibnamefont{and}
  \bibinfo{author}{\bibfnamefont{N.}~\bibnamefont{Triantafyllidis}},
  \bibinfo{journal}{J. Mech. Phys. Solids} \textbf{\bibinfo{volume}{60}},
  \bibinfo{pages}{120} (\bibinfo{year}{2012}).

\bibitem[{\citenamefont{Zubarev}(2013)}]{zubarev2013effect}
\bibinfo{author}{\bibfnamefont{A.~Y.} \bibnamefont{Zubarev}},
  \bibinfo{journal}{Soft Matter} \textbf{\bibinfo{volume}{9}},
  \bibinfo{pages}{4985} (\bibinfo{year}{2013}).

\bibitem[{\citenamefont{Jarkova et~al.}(2003)\citenamefont{Jarkova, Pleiner,
  M\"uller, and Brand}}]{jarkova2003hydrodynamics}
\bibinfo{author}{\bibfnamefont{E.}~\bibnamefont{Jarkova}},
  \bibinfo{author}{\bibfnamefont{H.}~\bibnamefont{Pleiner}},
  \bibinfo{author}{\bibfnamefont{H.-W.} \bibnamefont{M\"uller}},
  \bibnamefont{and} \bibinfo{author}{\bibfnamefont{H.~R.} \bibnamefont{Brand}},
  \bibinfo{journal}{Phys. Rev. E} \textbf{\bibinfo{volume}{68}},
  \bibinfo{pages}{041706} (\bibinfo{year}{2003}).

\bibitem[{\citenamefont{Bohlius et~al.}(2004)\citenamefont{Bohlius, Brand, and
  Pleiner}}]{bohlius2004macroscopic}
\bibinfo{author}{\bibfnamefont{S.}~\bibnamefont{Bohlius}},
  \bibinfo{author}{\bibfnamefont{H.~R.} \bibnamefont{Brand}}, \bibnamefont{and}
  \bibinfo{author}{\bibfnamefont{H.}~\bibnamefont{Pleiner}},
  \bibinfo{journal}{Phys. Rev. E} \textbf{\bibinfo{volume}{70}},
  \bibinfo{pages}{061411} (\bibinfo{year}{2004}).

\bibitem[{\citenamefont{Weeber et~al.}(2012)\citenamefont{Weeber, Kantorovich,
  and Holm}}]{Weeber2012}
\bibinfo{author}{\bibfnamefont{R.}~\bibnamefont{Weeber}},
  \bibinfo{author}{\bibfnamefont{S.}~\bibnamefont{Kantorovich}},
  \bibnamefont{and} \bibinfo{author}{\bibfnamefont{C.}~\bibnamefont{Holm}},
  \bibinfo{journal}{Soft Matter} \textbf{\bibinfo{volume}{8}},
  \bibinfo{pages}{9923} (\bibinfo{year}{2012}).

\bibitem[{\citenamefont{Weeber et~al.}(2015)\citenamefont{Weeber, Kantorovich,
  and Holm}}]{weeber2015ferrogels}
\bibinfo{author}{\bibfnamefont{R.}~\bibnamefont{Weeber}},
  \bibinfo{author}{\bibfnamefont{S.}~\bibnamefont{Kantorovich}},
  \bibnamefont{and} \bibinfo{author}{\bibfnamefont{C.}~\bibnamefont{Holm}},
  \bibinfo{journal}{J. Magn. Magn. Mater.} \textbf{\bibinfo{volume}{383}},
  \bibinfo{pages}{262} (\bibinfo{year}{2015}).

\bibitem[{\citenamefont{Ryzhkov et~al.}(2015)\citenamefont{Ryzhkov, Melenev,
  Holm, and Raikher}}]{ryzhkov2015coarse}
\bibinfo{author}{\bibfnamefont{A.~V.} \bibnamefont{Ryzhkov}},
  \bibinfo{author}{\bibfnamefont{P.~V.} \bibnamefont{Melenev}},
  \bibinfo{author}{\bibfnamefont{C.}~\bibnamefont{Holm}}, \bibnamefont{and}
  \bibinfo{author}{\bibfnamefont{Y.~L.} \bibnamefont{Raikher}},
  \bibinfo{journal}{J. Magn. Magn. Mater.} \textbf{\bibinfo{volume}{383}},
  \bibinfo{pages}{277} (\bibinfo{year}{2015}).

\bibitem[{\citenamefont{Menzel}(2014)}]{menzel2014bridging}
\bibinfo{author}{\bibfnamefont{A.~M.} \bibnamefont{Menzel}},
  \bibinfo{journal}{J. Chem. Phys.} \textbf{\bibinfo{volume}{141}},
  \bibinfo{pages}{194907} (\bibinfo{year}{2014}).

\bibitem[{\citenamefont{Ivaneyko et~al.}(2014)\citenamefont{Ivaneyko,
  Toshchevikov, Saphiannikova, and Heinrich}}]{Ivaneyko2014}
\bibinfo{author}{\bibfnamefont{D.}~\bibnamefont{Ivaneyko}},
  \bibinfo{author}{\bibfnamefont{V.}~\bibnamefont{Toshchevikov}},
  \bibinfo{author}{\bibfnamefont{M.}~\bibnamefont{Saphiannikova}},
  \bibnamefont{and} \bibinfo{author}{\bibfnamefont{G.}~\bibnamefont{Heinrich}},
  \bibinfo{journal}{Soft Matter} \textbf{\bibinfo{volume}{10}},
  \bibinfo{pages}{2213} (\bibinfo{year}{2014}).

\bibitem[{\citenamefont{Pessot et~al.}(2015)\citenamefont{Pessot, Weeber, Holm,
  L{\"o}wen, and Menzel}}]{pessot2015towards}
\bibinfo{author}{\bibfnamefont{G.}~\bibnamefont{Pessot}},
  \bibinfo{author}{\bibfnamefont{R.}~\bibnamefont{Weeber}},
  \bibinfo{author}{\bibfnamefont{C.}~\bibnamefont{Holm}},
  \bibinfo{author}{\bibfnamefont{H.}~\bibnamefont{L{\"o}wen}},
  \bibnamefont{and} \bibinfo{author}{\bibfnamefont{A.~M.}
  \bibnamefont{Menzel}}, \bibinfo{journal}{arXiv preprint, arXiv:1502.03707}
  (\bibinfo{year}{2015}).

\bibitem[{\citenamefont{Roeder et~al.}(2014)\citenamefont{Roeder,
  Reckenth\"aler, Belkoura, Roitsch, Strey, and Schmidt}}]{Roeder2014}
\bibinfo{author}{\bibfnamefont{L.}~\bibnamefont{Roeder}},
  \bibinfo{author}{\bibfnamefont{M.}~\bibnamefont{Reckenth\"aler}},
  \bibinfo{author}{\bibfnamefont{L.}~\bibnamefont{Belkoura}},
  \bibinfo{author}{\bibfnamefont{S.}~\bibnamefont{Roitsch}},
  \bibinfo{author}{\bibfnamefont{R.}~\bibnamefont{Strey}}, \bibnamefont{and}
  \bibinfo{author}{\bibfnamefont{A.~M.} \bibnamefont{Schmidt}},
  \bibinfo{journal}{Macromolecules} \textbf{\bibinfo{volume}{47}},
  \bibinfo{pages}{7200} (\bibinfo{year}{2014}).

\bibitem[{\citenamefont{Frickel et~al.}(2009)\citenamefont{Frickel, Messing,
  Gelbrich, and Schmidt}}]{frickel2009functional}
\bibinfo{author}{\bibfnamefont{N.}~\bibnamefont{Frickel}},
  \bibinfo{author}{\bibfnamefont{R.}~\bibnamefont{Messing}},
  \bibinfo{author}{\bibfnamefont{T.}~\bibnamefont{Gelbrich}}, \bibnamefont{and}
  \bibinfo{author}{\bibfnamefont{A.~M.} \bibnamefont{Schmidt}},
  \bibinfo{journal}{Langmuir} \textbf{\bibinfo{volume}{26}},
  \bibinfo{pages}{2839} (\bibinfo{year}{2009}).

\bibitem[{\citenamefont{Frickel et~al.}(2011)\citenamefont{Frickel, Messing,
  and Schmidt}}]{frickel2011magneto}
\bibinfo{author}{\bibfnamefont{N.}~\bibnamefont{Frickel}},
  \bibinfo{author}{\bibfnamefont{R.}~\bibnamefont{Messing}}, \bibnamefont{and}
  \bibinfo{author}{\bibfnamefont{A.~M.} \bibnamefont{Schmidt}},
  \bibinfo{journal}{J. Mater. Chem.} \textbf{\bibinfo{volume}{21}},
  \bibinfo{pages}{8466} (\bibinfo{year}{2011}).

\bibitem[{\citenamefont{Messing et~al.}(2011)\citenamefont{Messing, Frickel,
  Belkoura, Strey, Rahn, Odenbach, and Schmidt}}]{messing2011cobalt}
\bibinfo{author}{\bibfnamefont{R.}~\bibnamefont{Messing}},
  \bibinfo{author}{\bibfnamefont{N.}~\bibnamefont{Frickel}},
  \bibinfo{author}{\bibfnamefont{L.}~\bibnamefont{Belkoura}},
  \bibinfo{author}{\bibfnamefont{R.}~\bibnamefont{Strey}},
  \bibinfo{author}{\bibfnamefont{H.}~\bibnamefont{Rahn}},
  \bibinfo{author}{\bibfnamefont{S.}~\bibnamefont{Odenbach}}, \bibnamefont{and}
  \bibinfo{author}{\bibfnamefont{A.~M.} \bibnamefont{Schmidt}},
  \bibinfo{journal}{Macromolecules} \textbf{\bibinfo{volume}{44}},
  \bibinfo{pages}{2990} (\bibinfo{year}{2011}).

\bibitem[{\citenamefont{Csetneki et~al.}(2006)\citenamefont{Csetneki,
  Filipcsei, and Zr\'inyi}}]{Csetneki2006}
\bibinfo{author}{\bibfnamefont{I.}~\bibnamefont{Csetneki}},
  \bibinfo{author}{\bibfnamefont{G.}~\bibnamefont{Filipcsei}},
  \bibnamefont{and} \bibinfo{author}{\bibfnamefont{M.}~\bibnamefont{Zr\'inyi}},
  \bibinfo{journal}{Macromolecules} \textbf{\bibinfo{volume}{39}},
  \bibinfo{pages}{1939} (\bibinfo{year}{2006}).

\bibitem[{\citenamefont{An et~al.}(2014)\citenamefont{An, Groenewold, Picken,
  and Mendes}}]{An2014}
\bibinfo{author}{\bibfnamefont{H.~N.} \bibnamefont{An}},
  \bibinfo{author}{\bibfnamefont{J.}~\bibnamefont{Groenewold}},
  \bibinfo{author}{\bibfnamefont{S.~J.} \bibnamefont{Picken}},
  \bibnamefont{and} \bibinfo{author}{\bibfnamefont{E.}~\bibnamefont{Mendes}},
  \bibinfo{journal}{Soft Matter} \textbf{\bibinfo{volume}{10}},
  \bibinfo{pages}{997} (\bibinfo{year}{2014}).

\bibitem[{\citenamefont{Minsky}(1988)}]{Minsky1988}
\bibinfo{author}{\bibfnamefont{M.}~\bibnamefont{Minsky}},
  \bibinfo{journal}{Scanning} \textbf{\bibinfo{volume}{10}},
  \bibinfo{pages}{128} (\bibinfo{year}{1988}).

\bibitem[{\citenamefont{Roth et~al.}(2011)\citenamefont{Roth, Franzmann,
  D'Acunzi, Kreiter, and Auernhammer}}]{roth2011}
\bibinfo{author}{\bibfnamefont{M.}~\bibnamefont{Roth}},
  \bibinfo{author}{\bibfnamefont{M.}~\bibnamefont{Franzmann}},
  \bibinfo{author}{\bibfnamefont{M.}~\bibnamefont{D'Acunzi}},
  \bibinfo{author}{\bibfnamefont{M.}~\bibnamefont{Kreiter}}, \bibnamefont{and}
  \bibinfo{author}{\bibfnamefont{G.~K.} \bibnamefont{Auernhammer}},
  \bibinfo{journal}{arXiv preprint, arXiv:1106.3623}  (\bibinfo{year}{2011}).

\bibitem[{\citenamefont{Roth et~al.}(2012)\citenamefont{Roth, Schilde, Lellig,
  Kwade, and Auernhammer}}]{Roth2012}
\bibinfo{author}{\bibfnamefont{M.}~\bibnamefont{Roth}},
  \bibinfo{author}{\bibfnamefont{C.}~\bibnamefont{Schilde}},
  \bibinfo{author}{\bibfnamefont{P.}~\bibnamefont{Lellig}},
  \bibinfo{author}{\bibfnamefont{A.}~\bibnamefont{Kwade}}, \bibnamefont{and}
  \bibinfo{author}{\bibfnamefont{G.~K.} \bibnamefont{Auernhammer}},
  \bibinfo{journal}{Eur. Phys. J. E.} \textbf{\bibinfo{volume}{35}},
  \bibinfo{pages}{124} (\bibinfo{year}{2012}).

\bibitem[{\citenamefont{Wilhelm}(2008)}]{Wilhelm2008}
\bibinfo{author}{\bibfnamefont{C.}~\bibnamefont{Wilhelm}},
  \bibinfo{journal}{Phys. Rev. Lett.} \textbf{\bibinfo{volume}{101}},
  \bibinfo{pages}{028101} (\bibinfo{year}{2008}).

\bibitem[{\citenamefont{K\"ornig et~al.}(2014)\citenamefont{K\"ornig, Dong,
  Bennet, Widdrat, Andert, M\"uller, Sch\"uler, Klumpp, and
  Faivre}}]{Kornig2014}
\bibinfo{author}{\bibfnamefont{A.}~\bibnamefont{K\"ornig}},
  \bibinfo{author}{\bibfnamefont{J.}~\bibnamefont{Dong}},
  \bibinfo{author}{\bibfnamefont{M.}~\bibnamefont{Bennet}},
  \bibinfo{author}{\bibfnamefont{M.}~\bibnamefont{Widdrat}},
  \bibinfo{author}{\bibfnamefont{J.}~\bibnamefont{Andert}},
  \bibinfo{author}{\bibfnamefont{F.~D.} \bibnamefont{M\"uller}},
  \bibinfo{author}{\bibfnamefont{D.}~\bibnamefont{Sch\"uler}},
  \bibinfo{author}{\bibfnamefont{S.}~\bibnamefont{Klumpp}}, \bibnamefont{and}
  \bibinfo{author}{\bibfnamefont{D.}~\bibnamefont{Faivre}},
  \bibinfo{journal}{Nano Lett.} \textbf{\bibinfo{volume}{14}},
  \bibinfo{pages}{4653} (\bibinfo{year}{2014}).

\bibitem[{sup()}]{suppl}
\bibinfo{note}{See supplementary material.}

\bibitem[{\citenamefont{Mason et~al.}(2000)\citenamefont{Mason, Gisler, Kroy,
  Frey, and Weitz}}]{Mason2000}
\bibinfo{author}{\bibfnamefont{T.~G.} \bibnamefont{Mason}},
  \bibinfo{author}{\bibfnamefont{T.}~\bibnamefont{Gisler}},
  \bibinfo{author}{\bibfnamefont{K.}~\bibnamefont{Kroy}},
  \bibinfo{author}{\bibfnamefont{E.}~\bibnamefont{Frey}}, \bibnamefont{and}
  \bibinfo{author}{\bibfnamefont{D.~A.} \bibnamefont{Weitz}},
  \bibinfo{journal}{J. Rheol.} \textbf{\bibinfo{volume}{44}},
  \bibinfo{pages}{917} (\bibinfo{year}{2000}).

\bibitem[{\citenamefont{Raich and Bl\"umler}(2004)}]{Raich2004}
\bibinfo{author}{\bibfnamefont{H.}~\bibnamefont{Raich}} \bibnamefont{and}
  \bibinfo{author}{\bibfnamefont{P.}~\bibnamefont{Bl\"umler}},
  \bibinfo{journal}{Concept. Magn. Reson. B} \textbf{\bibinfo{volume}{23}},
  \bibinfo{pages}{16} (\bibinfo{year}{2004}).

\bibitem[{\citenamefont{Goubault et~al.}(2003)\citenamefont{Goubault, Jop,
  Fermigier, Baudry, Bertrand, and Bibette}}]{Goubault2003}
\bibinfo{author}{\bibfnamefont{C.}~\bibnamefont{Goubault}},
  \bibinfo{author}{\bibfnamefont{P.}~\bibnamefont{Jop}},
  \bibinfo{author}{\bibfnamefont{M.}~\bibnamefont{Fermigier}},
  \bibinfo{author}{\bibfnamefont{J.}~\bibnamefont{Baudry}},
  \bibinfo{author}{\bibfnamefont{E.}~\bibnamefont{Bertrand}}, \bibnamefont{and}
  \bibinfo{author}{\bibfnamefont{J.}~\bibnamefont{Bibette}},
  \bibinfo{journal}{Phys. Rev. Lett.} \textbf{\bibinfo{volume}{91}},
  \bibinfo{pages}{260802} (\bibinfo{year}{2003}).

\bibitem[{\citenamefont{Shcherbakov and Winklhofer}(2004)}]{Shcherbakov2004}
\bibinfo{author}{\bibfnamefont{V.~P.} \bibnamefont{Shcherbakov}}
  \bibnamefont{and}
  \bibinfo{author}{\bibfnamefont{M.}~\bibnamefont{Winklhofer}},
  \bibinfo{journal}{Phys. Rev. E} \textbf{\bibinfo{volume}{70}},
  \bibinfo{pages}{061803} (\bibinfo{year}{2004}).

\bibitem[{Ima()}]{Imagej}
\bibinfo{note}{\lowercase{h}ttp://imagej.nih.gov/ij/}.

\bibitem[{\citenamefont{Spieler et~al.}(2013)\citenamefont{Spieler,
  K{\"a}stner, Goldmann, Brummund, and Ulbricht}}]{spieler2013xfem}
\bibinfo{author}{\bibfnamefont{C.}~\bibnamefont{Spieler}},
  \bibinfo{author}{\bibfnamefont{M.}~\bibnamefont{K{\"a}stner}},
  \bibinfo{author}{\bibfnamefont{J.}~\bibnamefont{Goldmann}},
  \bibinfo{author}{\bibfnamefont{J.}~\bibnamefont{Brummund}}, \bibnamefont{and}
  \bibinfo{author}{\bibfnamefont{V.}~\bibnamefont{Ulbricht}},
  \bibinfo{journal}{Acta Mech.} \textbf{\bibinfo{volume}{224}},
  \bibinfo{pages}{2453} (\bibinfo{year}{2013}).

\bibitem[{\citenamefont{Sbalzarini and Koumoutsakos}(2005)}]{Sbalzarini2005}
\bibinfo{author}{\bibfnamefont{I.~F.} \bibnamefont{Sbalzarini}}
  \bibnamefont{and}
  \bibinfo{author}{\bibfnamefont{P.}~\bibnamefont{Koumoutsakos}},
  \bibinfo{journal}{J. Struct. Biol.} \textbf{\bibinfo{volume}{151}},
  \bibinfo{pages}{182} (\bibinfo{year}{2005}).

\bibitem[{\citenamefont{Annunziata et~al.}(2013)\citenamefont{Annunziata,
  Menzel, and L{\"o}wen}}]{annunziata2013hardening}
\bibinfo{author}{\bibfnamefont{M.~A.} \bibnamefont{Annunziata}},
  \bibinfo{author}{\bibfnamefont{A.~M.} \bibnamefont{Menzel}},
  \bibnamefont{and}
  \bibinfo{author}{\bibfnamefont{H.}~\bibnamefont{L{\"o}wen}},
  \bibinfo{journal}{J. Chem. Phys.} \textbf{\bibinfo{volume}{138}},
  \bibinfo{pages}{204906} (\bibinfo{year}{2013}).

\bibitem[{\citenamefont{Prokopieva et~al.}(2009)\citenamefont{Prokopieva,
  Danilov, Kantorovich, and Holm}}]{prokopieva09a}
\bibinfo{author}{\bibfnamefont{T.}~\bibnamefont{Prokopieva}},
  \bibinfo{author}{\bibfnamefont{V.}~\bibnamefont{Danilov}},
  \bibinfo{author}{\bibfnamefont{S.}~\bibnamefont{Kantorovich}},
  \bibnamefont{and} \bibinfo{author}{\bibfnamefont{C.}~\bibnamefont{Holm}},
  \bibinfo{journal}{Phys. Rev. E} \textbf{\bibinfo{volume}{80}},
  \bibinfo{pages}{031404} (\bibinfo{year}{2009}).

\bibitem[{\citenamefont{Tarama et~al.}(2014)\citenamefont{Tarama, Cremer,
  Borin, Odenbach, L{\"o}wen, and Menzel}}]{tarama2014tunable}
\bibinfo{author}{\bibfnamefont{M.}~\bibnamefont{Tarama}},
  \bibinfo{author}{\bibfnamefont{P.}~\bibnamefont{Cremer}},
  \bibinfo{author}{\bibfnamefont{D.~Y.} \bibnamefont{Borin}},
  \bibinfo{author}{\bibfnamefont{S.}~\bibnamefont{Odenbach}},
  \bibinfo{author}{\bibfnamefont{H.}~\bibnamefont{L{\"o}wen}},
  \bibnamefont{and} \bibinfo{author}{\bibfnamefont{A.~M.}
  \bibnamefont{Menzel}}, \bibinfo{journal}{Phys. Rev. E}
  \textbf{\bibinfo{volume}{90}}, \bibinfo{pages}{042311}
  (\bibinfo{year}{2014}).

\bibitem[{\citenamefont{Limbach et~al.}(2006)\citenamefont{Limbach, Arnold,
  Mann, and Holm}}]{limbach06a}
\bibinfo{author}{\bibfnamefont{H.~J.} \bibnamefont{Limbach}},
  \bibinfo{author}{\bibfnamefont{A.}~\bibnamefont{Arnold}},
  \bibinfo{author}{\bibfnamefont{B.~A.} \bibnamefont{Mann}}, \bibnamefont{and}
  \bibinfo{author}{\bibfnamefont{C.}~\bibnamefont{Holm}},
  \bibinfo{journal}{Comp. Phys. Comm.} \textbf{\bibinfo{volume}{174}},
  \bibinfo{pages}{704} (\bibinfo{year}{2006}).

\bibitem[{\citenamefont{Arnold et~al.}(2013)\citenamefont{Arnold, Lenz,
  Kesselheim, Weeber, Fahrenberger, R{\"o}hm, Ko\v{s}ovan, and
  Holm}}]{arnold13a}
\bibinfo{author}{\bibfnamefont{A.}~\bibnamefont{Arnold}},
  \bibinfo{author}{\bibfnamefont{O.}~\bibnamefont{Lenz}},
  \bibinfo{author}{\bibfnamefont{S.}~\bibnamefont{Kesselheim}},
  \bibinfo{author}{\bibfnamefont{R.}~\bibnamefont{Weeber}},
  \bibinfo{author}{\bibfnamefont{F.}~\bibnamefont{Fahrenberger}},
  \bibinfo{author}{\bibfnamefont{D.}~\bibnamefont{R{\"o}hm}},
  \bibinfo{author}{\bibfnamefont{P.}~\bibnamefont{Ko\v{s}ovan}},
  \bibnamefont{and} \bibinfo{author}{\bibfnamefont{C.}~\bibnamefont{Holm}}, in
  \emph{\bibinfo{booktitle}{Meshfree Methods for Partial Differential Equations
  {VI}}}, edited by \bibinfo{editor}{\bibfnamefont{M.}~\bibnamefont{Griebel}}
  \bibnamefont{and} \bibinfo{editor}{\bibfnamefont{M.~A.}
  \bibnamefont{Schweitzer}} (\bibinfo{publisher}{Springer},
  \bibinfo{year}{2013}), vol.~\bibinfo{volume}{89} of
  \emph{\bibinfo{series}{Lecture Notes in Computational Science and
  Engineering}}, pp. \bibinfo{pages}{1--23}.

\bibitem[{\citenamefont{Warner}(1972)}]{warner1972kinetic}
\bibinfo{author}{\bibfnamefont{H.~R.} \bibnamefont{Warner}},
  \bibinfo{journal}{Ind. Eng. Chem. Fundam.} \textbf{\bibinfo{volume}{11}},
  \bibinfo{pages}{379} (\bibinfo{year}{1972}).

\bibitem[{\citenamefont{Weeks et~al.}(1971)\citenamefont{Weeks, Chandler, and
  Andersen}}]{weeks71a}
\bibinfo{author}{\bibfnamefont{J.~D.} \bibnamefont{Weeks}},
  \bibinfo{author}{\bibfnamefont{D.}~\bibnamefont{Chandler}}, \bibnamefont{and}
  \bibinfo{author}{\bibfnamefont{H.~C.} \bibnamefont{Andersen}},
  \bibinfo{journal}{J. Chem. Phys.} \textbf{\bibinfo{volume}{54}},
  \bibinfo{pages}{5237} (\bibinfo{year}{1971}).

\bibitem[{\citenamefont{Fleischmann et~al.}(2012)\citenamefont{Fleischmann,
  Liang, Kapernaum, Giesselmann, Lagerwall, and Zentel}}]{fleischmann2012one}
\bibinfo{author}{\bibfnamefont{E.-K.} \bibnamefont{Fleischmann}},
  \bibinfo{author}{\bibfnamefont{H.-L.} \bibnamefont{Liang}},
  \bibinfo{author}{\bibfnamefont{N.}~\bibnamefont{Kapernaum}},
  \bibinfo{author}{\bibfnamefont{F.}~\bibnamefont{Giesselmann}},
  \bibinfo{author}{\bibfnamefont{J.}~\bibnamefont{Lagerwall}},
  \bibnamefont{and} \bibinfo{author}{\bibfnamefont{R.}~\bibnamefont{Zentel}},
  \bibinfo{journal}{Nature Commun.} \textbf{\bibinfo{volume}{3}},
  \bibinfo{pages}{1178} (\bibinfo{year}{2012}).

\end{thebibliography}

\end{document}